\documentclass[aip,amsmath,amssymb,reprint]{revtex4-1}

\usepackage[dvipsnames]{xcolor}
\usepackage{graphicx}
\usepackage{dcolumn}
\usepackage{bm}
\usepackage{comment}

\usepackage[utf8]{inputenc}
\usepackage[T1]{fontenc}
\usepackage{mathptmx}
\usepackage{etoolbox}
\usepackage{overpic}
\usepackage{DejaVuSans}
\usepackage{float}
\usepackage{adjustbox}
\usepackage[colorlinks=true, linkcolor=blue, citecolor=blue, urlcolor=blue]{hyperref}

\makeatletter
\def\@email#1#2{%
 \endgroup
 \patchcmd{\titleblock@produce}
  {\frontmatter@RRAPformat}
  {\frontmatter@RRAPformat{\produce@RRAP{*#1\href{mailto:#2}{#2}}}\frontmatter@RRAPformat}
  {}{}
}%
\makeatother
\begin{document}

\preprint{AIP/123-QED}

\title[]{Self-assembly and time-dependent control of active and passive triblock Janus colloids}

\author{Juri Franz Schubert}

\author{Salman Fariz Navas}%
\author{Sabine H. L. Klapp}
\email[]{sabine.klapp@tu-berlin.de}
\email[]{juri.schubert@tu-berlin.de}
\affiliation{ 
Institut für Physik und Astronomie, Technische Universität Berlin, Hardenbergstr. 36, 10623 Berlin,
Germany
}%

\date{\today}

\begin{abstract}

We perform Brownian Dynamics (BD) simulations to explore the self-assembly of a two-dimensional model system of triblock Janus colloids as an example of "patchy" colloids forming complex structures.
Previous experiments and simulation studies have shown that such systems are capable of forming a two-dimensional Kagome lattice at room temperatures. However, it is well established that the crystallization  is strongly hampered by the formation of long-living metastable aggregates. For this reason, recent studies have investigated {\em activity}, i.e., self-propulsion of the Janus particles as a mechanism to accelerate the formation of stable Kagome structures [Mallory and Cacciuto, JACS 141, 2500-2507 (2019)] at selected state points. Here we extend, first, the investigations of active Janus colloids for a broader range of densities and temperatures. We also characterize in detail the associated nucleation of Kagome clusters, as well as their structure in the steady state. Second, to make contact to the equilibrium case, we propose a simple activity time protocol where an initially chosen activity is switched off after a finite time. With this protocol, we not only find Kagome structures in a much broader range of densities than in the purely passive case, but also obtain a Kagome crystallization boundary very close to that proposed in earlier Monte Carlo simulations.

\end{abstract}

\maketitle

\section{\label{sec:level1}Introduction}

Self-assembly is a process where initially disordered particles or molecules aggregate to form larger ordered structures or even extended lattices.\cite{self-assembly, self-assembly2} This occurs spontaneously as a consequence of specific, often highly directional, interactions overcoming the omnipresent thermal fluctuations. The associated structure formation process has been attracting significant interest in the soft matter community for decades.\cite{sacanna1, kraft2012, sacanna2, PatchyAssemblyProgress} On the one hand, self-assembled structures are often promising from a material science point of view, an example being catalysis\cite{Catalysis1, Catalysis2022} or the development of novel optical materials\cite{PC1, PC2, Photonic2022, Photonic2023}. On the other hand, the phenomenon itself and, in particular, the pathway of self-assembly in strongly correlated particle systems raises a number of fundamental questions that are potentially relevant also in adjacent areas such as biophysics.\cite{VirusAssembly, VirusCapsid, HepBAssembly, VirusReviewNew}

In equilibrium (or "passive") self-assembly, the final structure corresponds to the global minimum of the system's free energy. It is well established that the pathway towards this final state is often hampered by high free energy barriers between metastable intermediate states, yielding extremely large equilibration times in both, simulation studies and experiments. For that reason, there is currently strong interest in exploring nonequilibrium routes to self-assembly. Indeed, it has already been shown that various sources of nonequilibrium, such as external driving fields\cite{Bisker}, but also intrinsic energy sources such as activity\cite{Bolhuis} and nonreciprocity\cite{GolestanianKineticTraps, Liebchen, Salman}, can help a self-assembling system to overcome kinetic barriers in complex free energy landscapes.
One main question of the present study is whether already a {\em finite} range of time in which the system is exposed to a nonequilibrium situation is sufficient to accelerate the overall assembly process.

We consider, as an example,  the two-dimensional (2D) self-assembly of spherical triblock Janus colloids. These are particles with two attractive "patches" at the poles (that, microscopically, arise from hydrophobic interactions in aqueous environments). As shown experimentally\cite{Granick} and in various simulation studies\cite{Sciortino, eslami2018,eslami2019,eslami2021,bahri2022}, triblock Janus colloids are able to self-assemble into an open crystal with Kagome structure, where the conventional hexagonal structure is diluted with vacant sites. Such open crystals are interesting candidates for photonic \cite{joannopoulos_openlattice, TriblockOpenPCRef} and porous media applications\cite{PorousMaterials}. For the specific case of triblock Janus colloids, the preferred formation of a Kagome lattice (in favour of a close-packed hexagonal lattice) can be explained by entropic arguments taking into account both translational and rotational vibrations.\cite{TPEntropy, CatesEntropy}
While the equilibrium self-assembly and phase diagram has been determined already some time ago in both, 2D\cite{Sciortino, eslami2018, bahri2022} and 3D \cite{Sciortino3D, 3DReinhart, reinhart_TJcrystal, 3DRao, 3DEslami}, "brute-force" Molecular Dynamics (MD) or Brownian Dynamics (BD) simulations often fail to reach the global minimum of the free energy. In particular, instead of observing Kagome structures, such simulations often predict more open web-like networks\cite{mallory2019} as long-lived intermediate states. These trapping phenomena are unfortunate as MD or BD simulations (or related methods) seem to be the most natural route to explore dynamical aspects of the aggregation process.  Thereby motivated, Mallory \textit{et al}. \cite{mallory2019} have recently suggested to use activity, i.e., equipping the particles with a mechanism that leads to self-propulsion, as a tool to stabilize Kagome lattices. Indeed, with activity they found large fractions of Kagome-bonded particles under conditions where no such assembly was possible in the passive case. This was interpreted as a hint that properly adjusted activity can help to dissolve metastable equilibrium structures. \\
Here we perform BD simulations based on the model potential proposed in Ref.~\citenum{mallory2019}. 
The motivation of the present study is twofold. First, inspired by the results in Ref.~\citenum{mallory2019}, who considered selected state points primarily in the low-temperature regime, we here explore active triblock systems in a broader range of densities and temperatures. Particular attention is given to temperatures close to the Kagome "crystallization boundary" predicted in earlier Monte Carlo (MC) simulations\cite{Sciortino}. Within the active case, we also perform a detailed analysis of the nucleation of Kagome clusters from the fluid state, using the Direct Forward Flux Sampling\cite{DFFS} technique recently used to study phase separation of active Brownian particles\cite{FFS} with isotropic interactions. Finally, having obtained a microscopic understanding of Kagome formation at high temperature, we propose a time protocol where (constant) activity is present only for a time interval that allows for intrinsic relaxation. Such protocols can be realized, in principle, using optical fields\cite{TunableByLight,LightInducedReverse, ExpActiveDoping} or hydrogen peroxide fuel\cite{ChemPropulsion, ExpActiveAmphiphilic}. With the protocol, we find stable Kagome structures in a broad range of densities and temperatures, also for the passive system. The upper boundary of the resulting Kagome "phase" is in good quantitative agreement with data from MC simulations\cite{Sciortino}.

The rest of the paper is organized as follows. In Section~\ref{sec:level2}, we review the investigated model\cite{mallory2019} and describe our method of performing a structural analysis on the aggregates. Section~\ref{sec:level5} starts with a summary of the key results of previous simulation studies. Subsequently, we turn to the discussion of our state diagrams obtained for the passive system, the active system, and finally a system subject to an activity-time protocol. We then perform a detailed analysis of the nucleation and inner structure of Kagome clusters in active systems. Finally, we summarize our findings in Section~\ref{sec:Conclusion}.
The paper is completed by several Appendices addressing a model comparison and technical details of our structural analysis.

\section{\label{sec:level2}Model}
\subsubsection{\label{sec:simulation}Model and Simulation}

\begin{figure}[]
   \centering
   \includegraphics[width=.45\textwidth]{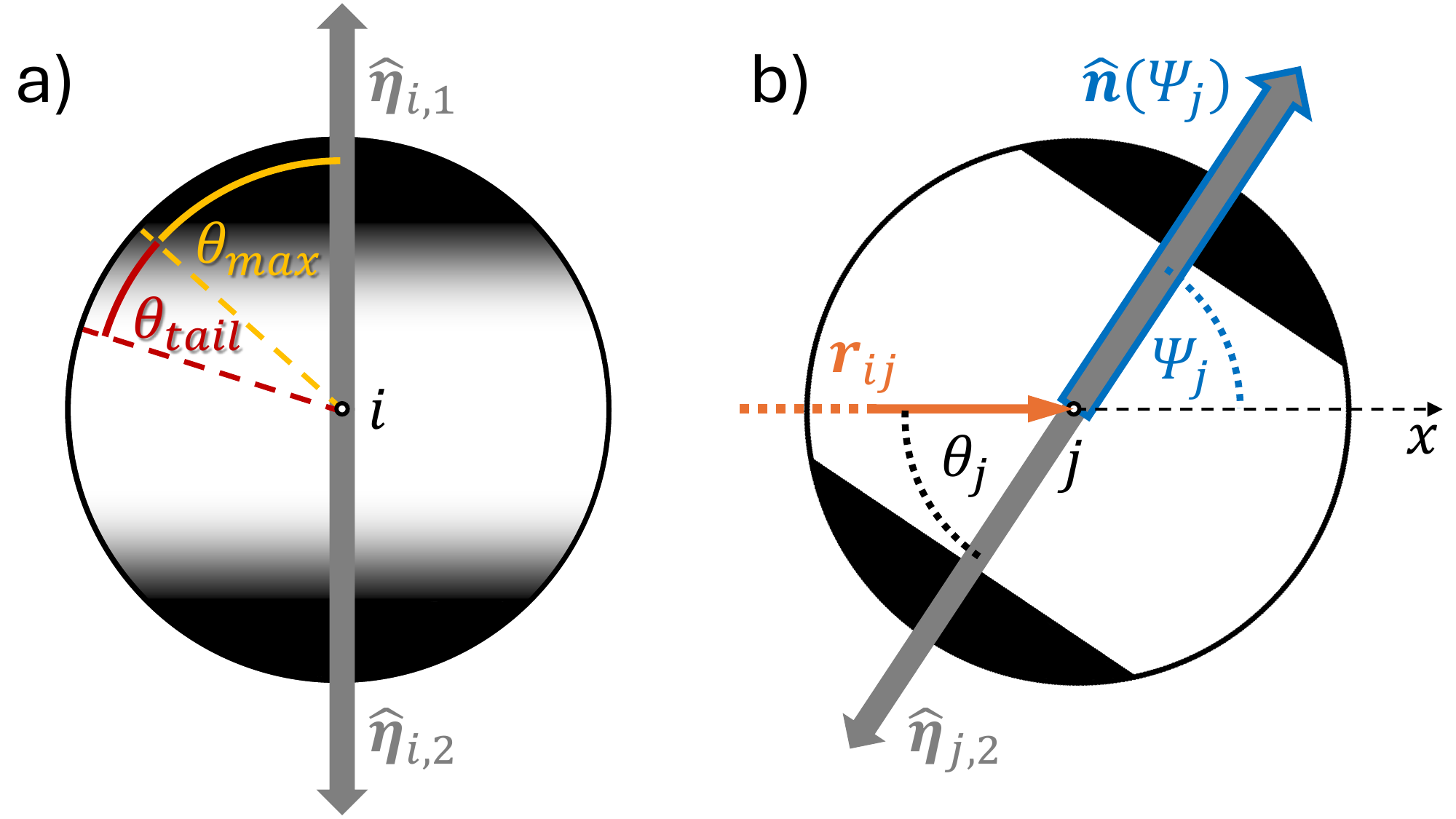}
    \caption[]{Schematic of a triblock Janus particle with repulsive equatorial region (white) and two attractive patches (black). 
    \textbf{a)} Definition of patch size and patch location. The latter are given via the antiparallel patch unit vectors $\hat{\pmb{\eta}}_{i,1}$ and $\hat{\pmb{\eta}}_{i,2}$ pointing towards the center of each patch. 
    The patch size is defined by the angles $\theta_{\text{max}}$ describing the region of constant angular potential (black), and $\theta_{\text{tail}}$ determining the region of decreasing angular potential (gradient of contrast), see Eq.~(\ref{eq:angular}). In our model, both patches have identical sizes. 
    \textbf{b)} Definition of the relevant angles. Particle orientation is given via the angle $\Psi$ between the heading vector $\hat{\pmb{n}}$ and the positive x-axis. $\theta_j$ is defined as the angle between the inter-particle vector $\pmb{r}_{ij}$, connecting $i$ and $j$, and one of the patch vectors $\hat{\pmb{\eta}}_{j}$. For simplicity, black patches are from here on represented solely by the region of constant angular potential, defined by $\theta_{\text{max}}$.}
    \label{fig:TriblockSchematic}
\end{figure}

Following an earlier study \cite{mallory2019}, we model the triblock Janus colloids as spherical particles of diameter $\sigma$ with two attractive patches (see Fig.~\ref{fig:TriblockSchematic}). The particles perform translational motion in the two-dimensional (2D) $xy$-plane and rotational motion within this plane.
The triblock nature of the particles is described by a repulsive equatorial region (white), that is located between two attractive patches (black) on opposing poles, as indicated in Fig.~\ref{fig:TriblockSchematic}. 
The center of the patch locations is characterized by two antiparallel patch unit vectors $\hat{\pmb{\eta}}_{i,1}$ and $\hat{\pmb{\eta}}_{i,2}$, one for each patch on particle $i=1,...,N$.
The patch size is described via the angles $\theta_{\text{max}}$ and $\theta_{\text{tail}}$, as indicated for a single patch in Fig.~\ref{fig:TriblockSchematic}a).
Specifically, $\theta_{\text{max}}$ controls a region of constant attraction around the center of the patch, while $\theta_{\text{tail}}$ controls the fading attractive interaction towards the repulsive region between $\theta_{\text{max}}$ and $\theta_{\text{max}}+\theta_{\text{tail}}$. In Fig.~\ref{fig:TriblockSchematic}a), the latter region is highlighted by a gradient of contrast to accentuate the smoothness of the underlying angular potential in this regime. In the following, the angle $\theta_{\text{max}}$ is used to set the size of both patches.\\
The position of particle $i$ is given by the 2D vector $\pmb{r}_i=(x_i,y_i)^T$, where $x_i$ and $y_i$ are cartesian coordinates.
The orientation of particle $i$ is described by the unit heading vector $\hat{\pmb{n}}(\psi_i)=(\cos\psi_i,  \sin\psi_i)^T$, which is pointing towards the center of one of the attractive patches. The polar angle $\psi_i$ is defined with respect to the positive x-direction. The two patch vectors are related to the unit heading vector via $\hat{\pmb{\eta}}_{i,1}=\hat{\pmb{n}}(\psi_i)$ and $\hat{\pmb{\eta}}_{i,2}=-\hat{\pmb{n}}(\psi_i)$, see Fig.~\ref{fig:TriblockSchematic}.\\
Two Janus particles $i$ and $j$ interact via the potential \cite{mallory2019}

\begin{equation}
    U_{ij}(\pmb{r}_{ij},\theta_i,\theta_j) = U_{\text{rep}}(r_{ij}) + U_{\text{att}}(r_{ij}) \phi(\theta_i) \phi(\theta_j) \quad ,  \label{eq:potential}
\end{equation}

where $\pmb{r}_{ij}=\pmb{r}_j - \pmb{r}_i$ is the connecting vector between the center of particles $i$ and $j$, $r_{ij}=|\pmb{r}_{ij}|$, and $\theta_i$ is the angle between the patch unit vector $\hat{\pmb{\eta}}_{i}$ and the interparticle vector $\pmb{r}_{ij}$ , as indicated in Fig.~\ref{fig:TriblockSchematic}b).
The angle $\theta_i$ (and $\theta_j$) is calculated from the scalar product 
of the normalized interparticle vector $\hat{\pmb{r}}_{ij} = \pmb{r}_{ij}/r_{ij}$ and the normalized patch unit vector $\hat{\pmb{\eta}}_{i}$, where $\hat{\pmb{\eta}}_{i}=\hat{\pmb{\eta}}_{i,1}(\psi_i)=\hat{\pmb{n}}(\psi_i)$ or $\hat{\pmb{\eta}}_{i}=\hat{\pmb{\eta}}_{i,2}(\psi_i)=-\hat{\pmb{n}}(\psi_i)$, depending on the patch that is engaged in the current interaction. 
The angle $\theta_i$ follows as $\theta_i=\arccos (\hat{\pmb{r}}_{ij} \cdot \hat{\pmb{\eta}}_i)$.
The interaction potential $U_{ij}(\pmb{r}_{ij},\theta_i,\theta_j)$ is the same as that used by Mallory and Cacciuto \cite{mallory2019}, who adapted a potential introduced earlier by Miller and Cacciuto \cite{miller}. 
The potential has been proposed to describe the behavior of patchy particles at high salt concentration of the solvent\cite{Granick}, where the effect of the electrostatic interaction becomes negligible and only excluded volume interactions are relevant repulsion effects. In general, this potential has been tailored to match the behavior of experimental patchy particle systems phenomenologically.\cite{miller, ExpJanus}\\
The first term on the right-hand side of Eq.~(\ref{eq:potential}) represents the volume exclusion modelled by the repulsive part of a Lennard-Jones (LJ) potential that is, $U_{\text{rep}}(r_{ij}) =4 \epsilon_{\text{rep}} ( \sigma/r_{ij} )^{12}$, where $\epsilon_{\text{rep}}$ is the repulsive energy. We cut off the potential at $r_c=1.5\sigma$.
The second term describes the orientation-dependent interaction between the attractive patches. Here, the distance-dependent function $U_{\text{att}}(r_{ij})$ is given by

\begin{equation}
    U_{\text{att}}(r_{ij}) =
\begin{cases}
    4 \epsilon_{\text{att}} \biggl [ D(r_{ij})^{12} -  D(r_{ij})^{6} \biggr ]    &,\quad  r_{ij} \leq 1.5\sigma\\ 
    0                                                                           &, \quad  r_{ij} > 1.5\sigma \quad,  \\ 
\end{cases} \label{eq:Uatt} 
\end{equation}

where

\begin{equation}
    D(r_{ij})=\frac{\sigma/2}{|r_{ij} - \sigma| +  2^{1/6}\cdot\sigma/2} \quad . 
\end{equation}

\begin{figure}[]
   \centering
   \includegraphics[width=.5\textwidth]{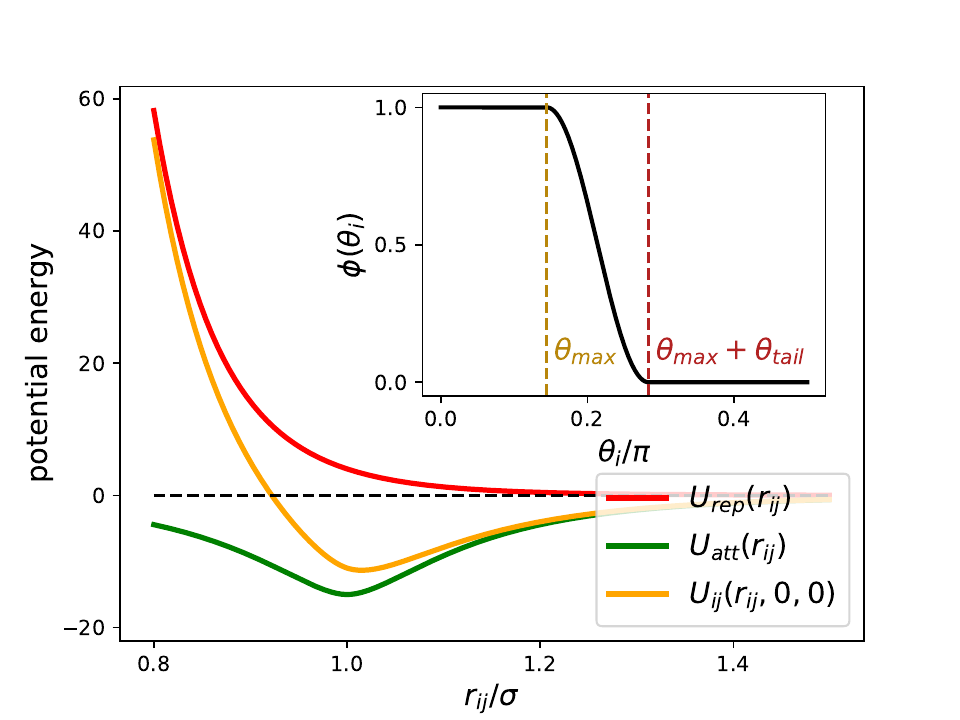}
    \caption[]{Interaction potential as function of distance $r_{ij}$ at fixed orientations $\theta_i=\theta_j=0$. The three colored curves show the purely repulsive potential $U_{\text{rep}}(r_{ij})$ in red, the purely attractive potential $U_{\text{att}}(r_{ij})$ in green, and the combined isotropic potential in yellow. The inset depicts the angular potential $\phi(\theta_i)$ according to Eq.~(\ref{eq:angular}), generated by keeping the orientation of particle $j$ fixed and rotating particle $i$ at constant inter-particle distance $r_{ij} = \sigma$. Parameters in Eq.~(\ref{eq:angular}) have been set to $\theta_{\text{max}} = 26^\circ$ and $\theta_{\text{max}} + \theta_{\text{tail}}=51^\circ$.}
    \label{fig:DistancePot}
\end{figure}

The potential $U_{\text{att}}(r_{ij})$ has a minimum with potential well depth $\epsilon_{\text{att}}=15\epsilon_{\text{rep}}$. This relation has been found \cite{mallory2019} to be most suitable for the formation of the Kagome lattice, while still being in the same order of magnitude as the experimental counterpart.\cite{Granick}\\
The orientational dependence of the patch-patch interaction in Eq.~(\ref{eq:potential}) is introduced through the functions $\phi(\theta_i)$ and $\phi(\theta_j)$, where

\begin{equation}
    \phi(\theta_i) = 
\begin{cases}
    1                                                                                   &, \quad \theta_i \leq \theta_{\text{max}}\\
    \cos^2 \Bigl (\frac{\pi(\theta_i-\theta_{\text{max}})}{2\theta_{\text{tail}}} \Bigr )             &, \quad \theta_{\text{max}} \leq \theta_i \leq \theta_{\text{max}}+\theta_{\text{tail}}\\
    0                                                                                   &, \quad \theta_i > \theta_{\text{max}}+\theta_{\text{tail}}   \quad .
\end{cases} \label{eq:angular}
\end{equation}

The function $\phi(\theta_j)$ follows analogously. Equation (\ref{eq:angular}) describes a smooth step function of the angle $\theta_i \in [0, \frac{\pi}{2}]$, as can be seen from the inset in  Fig.~\ref{fig:DistancePot}. The angles $\theta_{\text{max}}$ and $\theta_{\text{tail}}$ are explained in Fig.~\ref{fig:TriblockSchematic}a).

The translational and rotational dynamics of the system is described via the overdamped Langevin equations 

\begin{align}
    \gamma_t\dot{\pmb{r}}_i &= \sum_{j \neq i} \pmb{F}_{ij}(\pmb{r}_{ij},\theta_i,\theta_j) + F_0\hat{\pmb{n}}(\psi_i) + \sqrt{2k_BT\gamma_t} \ \pmb{\xi}_{t_i}(t) \label{eq:langevin1} \\
    \gamma_r \dot \psi_i &= \sum_{j \neq i} T_{ij}(\pmb{r}_{ij},\theta_i,\theta_j) + \sqrt{2k_BT\gamma_r} \ \xi_{r_i}(t) \quad , \label{eq:langevin2}
\end{align}

where the dots indicate first order derivatives in time, and $\gamma_t$ and $\gamma_r$ are friction constants of translational and rotational motion, respectively.
The first term on the right-hand side of Eq.~(\ref{eq:langevin1}) involves the pair forces $\pmb{F}_{ij}=-\partial_{\pmb{r}_i}U_{ij}$, where $U_{ij}$ is given in Eq.~(\ref{eq:potential}).
The second term accounts for the active nature of the triblock particles (if present), where the heading vector $\hat{\pmb{n}}(\psi_i)$ sets the direction of self-propulsion and $F_0$ the strength of the active force at which the particle is propelled. 
Following Ref.~\citenum{mallory2019}, the heading vector is chosen parallel to the particle axis and $F_0=\gamma_t v_0$, with $v_0$ being the self-propulsion velocity.
The third term in Eq.~(\ref{eq:langevin1}) describes the coupling of the particle to a heat bath, where $\pmb{\xi}_{t_{i}}(t)$ is a 2D translational Gaussian white noise with zero mean $\langle \xi_{t_{i, \nu}}(t) \rangle = 0$ and correlations $\langle \xi_{t_{i, \nu}}(t) \xi_{t_{j, \mu}}(t') \rangle = \delta_{ij}\delta_{\nu \mu}\delta(t-t')$, with $\nu, \mu \in \lbrace 1,\ 2 \rbrace$ being cartesian indices. The translational diffusion coefficient follows as $D_t = k_BT/\gamma_t$, where $k_BT$ is the thermal energy.\\
In the rotational equation of motion (\ref{eq:langevin2}), the first term on the right-hand side sums the individual torques $T_i= \pm|\hat{\pmb{\eta}}_i \times \partial_{\pmb{\hat{\eta}}_i}U_{ij}|$ on particle $i$, and $\xi_{r_{i}}(t)$ is a 1D white noise with zero mean $\langle \xi_{r_{i}}(t) \rangle = 0$ and correlation $\langle \xi_{r_{i}}(t) \xi_{r_{j}}(t') \rangle = \delta_{ij}\delta(t-t')$.\\

We perform BD simulations for systems of $N=1024$ and $N=2025$ triblock particles in a square simulation box with periodic boundaries. To non-dimensionalize the equations of motion, we use $k_BT$ as our energy scale, $\sigma$ as a length scale and the Brownian time $\tau_b=\sigma^2\gamma_t/k_BT$ as a time scale. The systems are then characterized by the dimensionless number density $\rho^*=\rho\sigma^2$, temperature $T^*=k_BT/\epsilon_{\text{rep}}$ and self-propulsion parameter $v^*=F_0\sigma/\epsilon_{\text{rep}}=T^*v_0\tau_b/\sigma$, which is referred to as the dimensionless propulsion velocity.
All simulations are initialized from a square lattice configuration of particles with random orientations.
The equations of motion Eq.~(\ref{eq:langevin1}) and Eq.~(\ref{eq:langevin2}) are integrated using the Euler-Maruyama method\cite{EulerScheme} for stochastic ordinary differential equations. The integration step size is chosen to be $10^{-5}\tau_b$ and simulations are carried out for $10^3\tau_b$ if not specified otherwise.

\subsubsection{\label{sec:StructureDetect}Structural analysis}

\begin{figure*}[]
    \centering
    \includegraphics[width=0.9\textwidth]{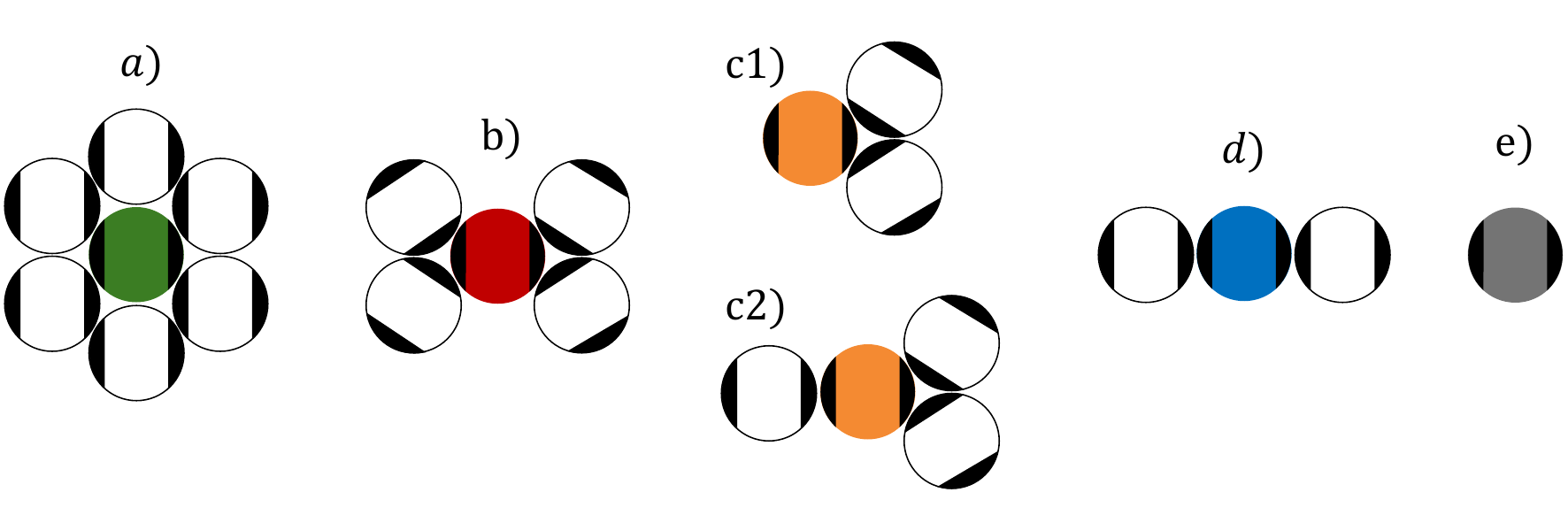}
    \caption{Schematics of (idealized) configurations occurring in the self-assembly of triblock Janus particles, (a) Hexagonal, (b) Kagome, (c1) and (c2) Triangular, (d) Chain-like and (e) Monomer.}
    \label{fig:structuretypes}
\end{figure*}

To characterize the system's structural properties during the self-assembly process and in the final state, we perform an (automatized) analysis of the configuration of each particle based on its immediate neighborhood.
Previous studies\cite{Granick, Sciortino} have shown that, within the investigated parameter regime, the following configurations occur: Monomer, Chain-like, Triangular, Kagome, and Hexagonal, as shown schematically in Fig.~\ref{fig:structuretypes}. To facilitate accurate structure detection during the simulation, we have developed an algorithm that evaluates particle neighbor lists by employing basic concepts of multiset theory. The algorithm returns unique numerical identifiers for each particle, that determine its specific local configuration. Here, two particles are considered bonded if $r_{ij}<1.15\sigma$ and an additional attractive interaction between any two of their patches is present, fulfilling the condition $\phi(\theta_i)\phi(\theta_j)>0$. Detailed explanations of the algorithm as well as formal definitions of configurations can be found in Appendix~\ref{sec:levelA2}.\\
For practical purposes, we group Monomer and Chain-like particles, including loose chain ends, into a larger class of Fluid-like particles. Examples for the identification of Kagome, Triangular, and Fluid-like particles based on simulation snapshots at a single time point are presented in Fig.~\ref{fig:three_images} (in this example the Hexagonal configuration does not occur). To determine the system's \textit{global} state from the particle-level structure information, the following procedure is employed. Each particle $i$ is characterized by a time-dependent structure type $s=s_i(t)=\{\text{Fluid-like, Triangular, Kagome, Hexagonal}\}$, evaluated for all configurations within the interval $t=t_0,\dots,t_n$ and stored in the list $S_i=[ s_i(t_0),\ s_i(t_1),\dots,\ s_i(t_n) ]$. Next, a time average is computed by taking the mode of $S_i$, i.e., $\langle S_i \rangle_t = \textit{Mode}(S_i)$, returning the most frequently observed structure type for particle $i$ within the selected time window. By summing over all particles that share the same average structure $\langle S_i \rangle_t = s$, we calculate the average structure fraction $f_s \in [0,1]$ that quantifies how prominently each structure contributes to the system’s overall state:

\begin{equation} f_s = \frac{1}{N}\sum_{\langle S_i \rangle_t = s} 1 \quad . \end{equation}

The time average is taken over $20\tau_b$. This time window is chosen such that at sufficiently high $T^*$, only the Fluid-like state remains and transient Triangular structures are averaged out. Based on the resulting fractions $f_s$ we then construct state diagrams, see Fig.~\ref{fig:PurelyPassiveDiagram} for an example.
Here, the values of $f_s$ at a given state point $(T^*,\rho^*)$ are indicated as color-coded segments on squares.

\section{\label{sec:level5}Results}
\subsection{\label{sec:level4}The Equilibrium Triblock Phase Diagram}

Several aspects of the phase behavior of the equilibrium triblock system have been investigated in previous simulation studies.\cite{Sciortino, eslami2018, eslami2019, eslami2021, mallory2019, bahri2022} In this section, we first summarize results which are particularly relevant for the present study. Subsequently, we present our own state diagram in Fig.~\ref{fig:PurelyPassiveDiagram} alongside with results that have been obtained in Ref.~\citenum{Sciortino} and Ref.~\citenum{bahri2022}. 

\subsubsection{\label{sec:Theory}Results from previous studies}

\begin{figure}[b]
   \centering
   \begin{overpic}[width=.5\textwidth]{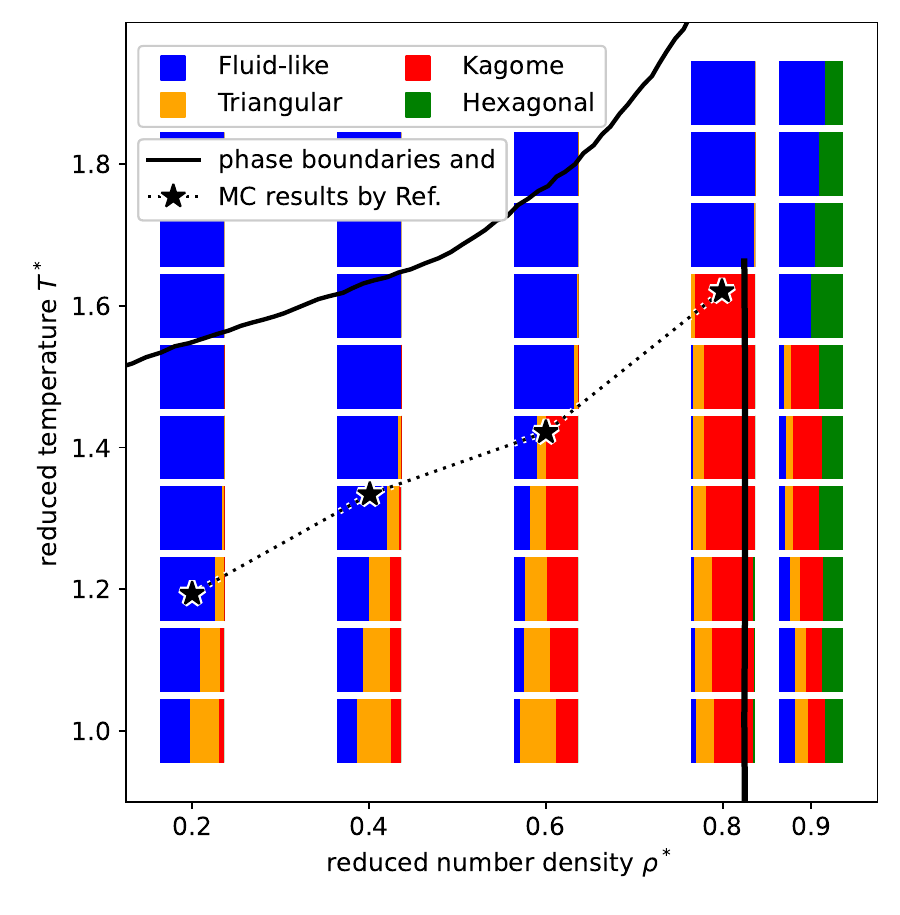}
   \put(49.1,76.04){\textsf{\fontsize{10}{14}\selectfont \cite{Sciortino}}}
   \end{overpic}
   \caption[]{$T^*$-$\rho^*$ state diagram of the equilibrium triblock system. The colored squares represent results of the present work. Specifically, each square corresponds to a $N=1024$ particle simulation, evaluated after $10^3\tau_b$ simulation time.
   The solid black lines have been obtained in Ref.~\citenum{Sciortino}. The monotonically increasing line corresponds to the upper branch of the Kagome-Fluid coexistence region. The vertical black line corresponds to the Solid-Solid (Kagome-Hexagonal) phase boundary. 
   Also shown are the Monte-Carlo (MC) results (black stars, connected by dotted line) representing the "crystallization boundary" (in the sense of spontaneous formation of solid clusters). All results from Ref.~\citenum{Sciortino} pertain to the Kern-Frenkel model with parameters as described in Appendix~\ref{sec:levelA1}.}
   \label{fig:PurelyPassiveDiagram}
\end{figure}

To our knowledge, the first study targeting the phase diagram of triblock Janus colloids was performed by Romano and Sciortino \cite{Sciortino} on the basis of a hard sphere Kern-Frenkel \cite{KFpotential} (KF) model. This model combines a short-range attractive square-well potential with a step-function-like angular potential describing the interaction between the particles' patches (for a summary, see Appendix~\ref{sec:levelA1}). In Ref.~\citenum{Sciortino}, the coexistence lines characterizing the first-order transition between fluid-like and solid-like (Kagome or Hexagonal) states were obtained using thermodynamic Gibbs-Duhem \cite{gibbs-duhem} integration. The resulting phase diagram for the patch width of $\theta_{\text{max}}\approx 33^\circ$ is shown in Fig.~\ref{fig:PurelyPassiveDiagram} (taken from Fig.~2(b) in Ref.~\citenum{Sciortino}). The upper branch of the Fluid-Solid coexistence line (obtained in Ref.~\citenum{Sciortino}) is given by the monotonically increasing solid black curve. For reduced temperatures $T^*$ above this line, the system is found in a fluid phase. Below the line, the overall structure depends on the density. A fully developed Kagome crystal, spanning the whole simulation box, occurs only within a narrow density region around the vertical black line at $\rho^*\approx 0.83$. The upper ($\rho^*\approx0.87$) and lower ($\rho^*\approx0.79$) bounds of this density range are determined by excluded-volume and the range of the interactions, respectively.\cite{Sciortino} For smaller densities, the Kagome crystal is only realized in coexistence with a dilute fluid phase.\cite{Sciortino}
To further study the crystallization behavior, the authors of Ref.~\citenum{Sciortino} performed separate Monte-Carlo (MC) simulations in the NVT ensemble for selected points in the $T^*$-$\rho^*$-plane. 
The black stars in Fig.~\ref{fig:PurelyPassiveDiagram} (connected by a dotted line) indicate the largest $T^*$ value where they observed spontaneous crystallization (Kagome or Hexagonal) from the fluid phase.
The shape of the so-obtained MC crystallization boundary roughly follows that of the Fluid-Solid coexistence line. However, the MC crystallization boundary is offset by $\Delta T^*\approx 0.4$ towards lower reduced temperatures. In addition, the predicted structures are different:
For densities $0.2 \leq \rho^* \leq 0.6$, MC simulations predict crystallization into the Kagome lattice, whereas $\rho^*=0.8$ yields a Hexagonal lattice.\cite{Sciortino}
We note that the authors of Ref.~\citenum{Sciortino} also searched for the occurrence of gas-liquid phase separation (driven by the overall attractive nature of the orientational interactions). For narrow patch widths such as the one used in the present study, however, crystallization typically occurs before any gas-liquid phase separation can be observed\cite{Sciortino, bahri2022}, i.e., the corresponding critical point is suppressed.\\
A further relevant study is that of Bahri \textit{et al}.\cite{bahri2022} who investigated a more detailed model of triblock particles involving soft sphere repulsion, an isotropic Yukawa potential (modelling screening), and an anisotropic LJ-like potential for the patch-patch interaction. 
Methodologically, the authors of Ref.~\citenum{bahri2022} used a form of metadynamics\cite{numericalbias, Metadynamics} simulations where biasing potentials are successively added to the interaction potential to drive the system between Fluid, Kagome and Hexagonal phases. The resulting phase diagram has a similar topology as that in Ref.~\citenum{Sciortino}. \\
An interesting aspect discussed in Ref.~\citenum{bahri2022} concerns the difficulty that comes with observing spontaneous crystallization of a Kagome crystal out of the metastable fluid state. Since the two states correspond to two distinct minima in the system's free energy, which are separated by a barrier, the authors of Ref.~\citenum{bahri2022} utilized the aforementioned metadynamics method to sample crystallization events within accessible simulation times.
A related problem arises under highly dilute conditions, especially at low $T^*$, resulting in the emergence of a metastable web-like state \cite{mallory2019} that is associated with kinetic arrest, which hinders the formation of a Kagome lattice. The latter behavior is addressed by Mallory and Cacciuto\cite{mallory2019}, who used BD simulations of the model introduced in Section \ref{sec:level2}. At low densities $\rho^*\approx 0.2$, they report the formation of a web-like network of particles with an overall fraction of Kagome particles close to zero. This effect becomes particularly relevant for $T^*{\scriptstyle \lessapprox} 1$, where thermal fluctuations are too weak to break malformed bonds. The most complete self-assembly into Kagome structures (i.e., largest value of $f_{\text{Kagome }}$) is found at $\rho^*\approx 0.8$.\cite{mallory2019}
To realize Kagome structures at smaller densities, the authors of Ref.~\citenum{mallory2019} proposed to choose an \textit{active} triblock system, where each particle self-propels in the direction of one of its patches at constant velocity. They showed that even weak self-propulsion leads to a significant increase of the overall fraction of Kagome particles, accompanied by a speedup of its formation.

\subsubsection{\label{sec:level6}State diagram in the purely passive case}

In the present study, we have extended the investigation of Ref.~\citenum{mallory2019}, where the focus was on low temperatures $T^* \leq 1.2$, towards higher temperatures in the range $2.0>T^* \geq 1.0$. Particular attention has been given to the states related to the Kagome crystallization boundary obtained in Ref.~\citenum{Sciortino} (see stars in Fig.~\ref{fig:PurelyPassiveDiagram}). In this way we obtained a full state diagram of the (passive) equilibrium system using unbiased BD simulations.
To enable a direct comparison with the Kern-Frenkel model used in Ref.~\citenum{Sciortino}, we establish a correspondence between the reduced temperatures $T^*$ of both models and set the patch opening angle to $\theta_{\max} = 26^\circ$. A detailed justification for this choice is given in Appendix~\ref{sec:levelA1}.\\
The resulting $T^*$-$\rho^*$ state diagram is presented in Fig.~\ref{fig:PurelyPassiveDiagram}, where each square represents results from a single simulation at the respective temperature and density. Each system was analyzed after a simulation time of $10^3\tau_b$ by detecting the relevant structure types via the algorithm presented in Appendix~\ref{sec:levelA2}. The colored fractions on each square in Fig.~\ref{fig:PurelyPassiveDiagram} directly correspond to the average fraction $f_s$ of particles belonging to one of the four structure types $s=\{\text{Fluid-like, Triangular, Kagome, Hexagonal} \}$, as defined in Section~\ref{sec:StructureDetect}.



We start the discussion from the high-density regime. At $\rho^*=0.9$ we find coexisting Fluid and Hexagonal structures for temperatures $T^*\geq 1.6$, with the fraction of Hexagonal particles decreasing as $T^*$ is increased. For $T^*<1.6$, the Fluid-Hexagonal coexistence is replaced by a three-phase coexistence of Kagome, Hexagonal and Fluid structures. These coexistence regions at $\rho^*=0.9$ were not specifically mentioned in Ref.~\citenum{Sciortino}, but were discussed in the later study Ref.~\citenum{bahri2022}, who qualitatively obtained the same phase behavior.
At $\rho^*=0.8$ the system's global density is slightly below the density of the fully developed Kagome lattice and we observe the formation of an almost ideal Kagome lattice at $T^* = 1.6$. This finding contrasts with Ref.~\citenum{Sciortino}, where crystallization into Hexagonal structures was reported at this density and temperature. The discrepancy can be attributed to the soft-sphere nature of the interaction potential Eq.~(\ref{eq:potential}) since, as suggested in Ref.~\citenum{bahri2022}, the potential softness favors self-assembly of the Kagome lattice over the Hexagonal one. 
At the lower density $\rho^*=0.6$, the Kagome lattice starts to appear in combination with Fluid-like structures, indicating Kagome-Fluid coexistence.
Overall, comparison to the MC results in Ref.~\citenum{Sciortino} (black stars) reveals that the present predictions of the onset of Kagome crystallization from the Fluid state, within the density range $0.6 \leq \rho^* \leq 0.8$, are in good agreement despite the differences between the underlying models. At lower densities, however, we rather observe a smooth transition from the purely Fluid-like state into a Triangular-dominated state with a marginally small fraction of Kagome particles (in contrast to Ref.~\citenum{Sciortino}).
We also note that there was no indication of an underlying gas-liquid separation.\\
We understand the Triangular structure as an intermediate structure that forms during the self-assembly into the final Kagome lattice. 
This interpretation is justified as follows. Across all densities, the fraction of triangular particles increases as $T^*$ is decreased. On the particle level, this trend corresponds to an increase of defects within the Kagome crystal (for $\rho^* = 0.8$), and an enhanced branching of particle chains in predominantly Fluid-like states (for $\rho^*\leq 0.4$). Such behavior has also been reported in Ref.~\citenum{mallory2019}.
Moreover, the resulting web-like configurations, which contain a large fraction of triangular structures, have also been observed in the experimental triblock system at $\rho^*\approx 0.8$. Here, they appear as intermediate structures that gradually transform into the energetically favored Kagome lattice over a timescale of approximately more than $50$ hours.\cite{Granick}\\
Coming back to the low-density regime in Fig.~\ref{fig:PurelyPassiveDiagram}, Ref.~\citenum{Sciortino} and Ref.~\citenum{bahri2022} suggest that the Fluid-Kagome coexistence region should extend into the region $\rho^*\leq0.4$, contrary to what we observe. In the following, we show how activity can be utilized to obtain a sharp Kagome-Fluid boundary across all densities considered.


\begin{figure}[]
   \centering
   \begin{overpic}[width=.5\textwidth]{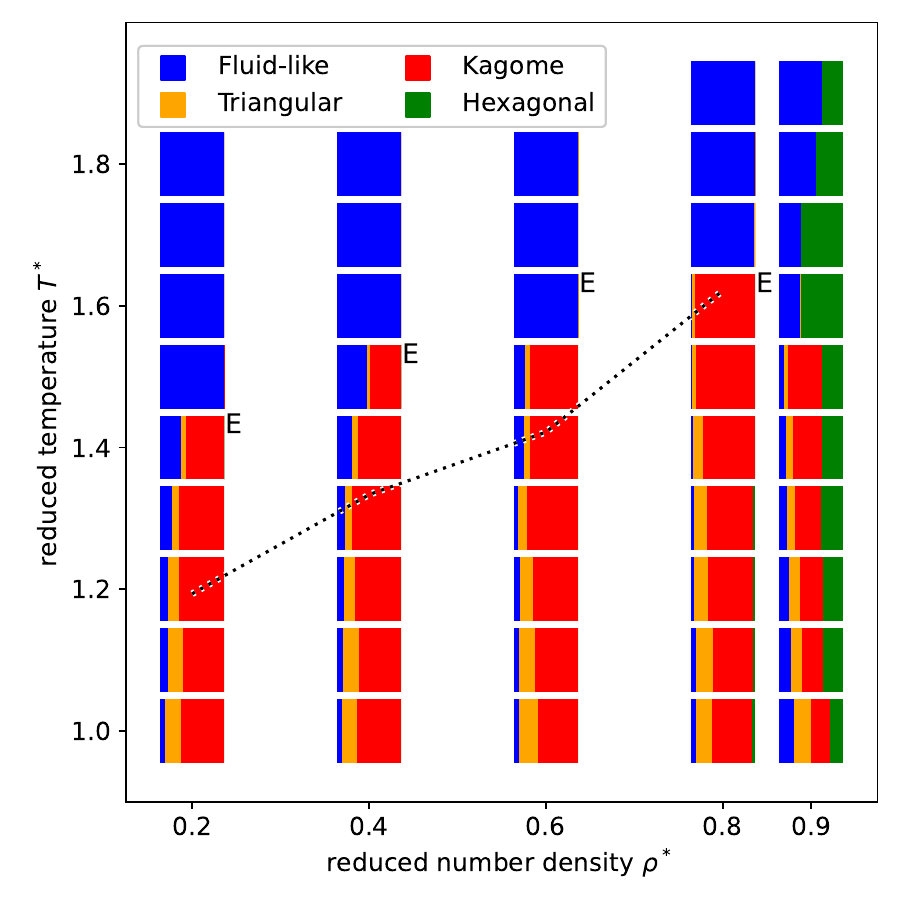}
   \end{overpic}
   \caption[]{$T^*$-$\rho^*$ state diagram of our active triblock system at propulsion velocity $v^*=4$, evaluated for the $N=1024$ particle system after $10^3\tau_b$. Squares marked with "E" to the top right indicate that a small ensemble of ten simulations was used to determine the state, due to the rare event character of crystallization near the Fluid-Kagome boundary. In such cases, only one among the successful simulations, if present at all, is analyzed after steady state is reached. The dotted line is kept as a reference to the "crystallization boundary" from Ref.~\citenum{Sciortino} in the passive system.}
   \label{fig:ActiveDiagram}
\end{figure}

\subsection{\label{sec:level7}Active state diagram and activity-time protocol sampling}
\subsubsection{\label{sec:level7}Active state diagram}
We now examine how the state diagram in Fig.~\ref{fig:PurelyPassiveDiagram} changes when, instead of the passive system characterized by $v^*=0$, an active system is considered. Specifically, we choose the reduced propulsion velocity $v^*=4$, while all remaining parameters remain unchanged.
At this value of $v^*$, Ref.~\citenum{mallory2019} reported the largest Kagome yields throughout the density range considered. Our results are shown in Fig.~\ref{fig:ActiveDiagram}.
For low temperatures and densities in the range $\rho\leq 0.6$, we find that the activity results in a larger fraction of Kagome particles as compared to the passive case. This is accompanied by a reduction of the Triangular and Fluid fractions. 
The most significant effect is found at $\rho^*=0.2$.
Upon increase of the temperature from small values, the data indicate a boundary, separating state points dominated by Kagome structures from those that remain Fluid-like. 
At this boundary, metastable fluid-like states emerge, leading to a low (yet finite) probability of observing crystallization into Kagome clusters (for further details, see Section~\ref{sec:level8}). To capture such potential crystallization events, we conduct an ensemble of ten independent simulations for each state point marked by the letter "E" in Fig.~\ref{fig:ActiveDiagram}. Each color-coded segment associated with an "E" state point represents results from a single simulation, evaluated in steady-state. In those cases, where the color code indicates a Kagome fraction, at least one out of the ten simulations resulted in successful crystallization.
At the so-obtained Kagome-Fluid boundary, Kagome structures generally form a large, stable cluster coexisting with a dilute Fluid-like phase, as illustrated in the exemplary simulation snapshot shown in the inset of Fig.~\ref{fig:Cluster_Stability}b) in Section~\ref{sec:level10}, where further analysis on the steady-state configuration of Kagome clusters is conducted.  \\
For higher densities ($\rho^* = 0.8$ and $\rho^* = 0.9$), the differences between the active and passive systems are minimal when we compare the structures at given $(T^*,\rho^*)$.
Notably, the Kagome region of the active state diagram in Fig.~\ref{fig:ActiveDiagram} fully encloses the range of densities and temperatures below the crystallization boundary found in Ref.~\citenum{Sciortino} (dotted line) for the passive system.\\
Overall, the topology of the active state diagram in Fig.~\ref{fig:ActiveDiagram} is quite similar to that of the passive, equilibrium case. These parallels already suggest that activity can be used as a tool to ease the sampling problems encountered in the passive case.

\subsubsection{\label{sec:level7}Improved passive state diagram via activity-time protocol sampling}

\begin{figure}[b]
   \centering
   \begin{overpic}[width=.5\textwidth]{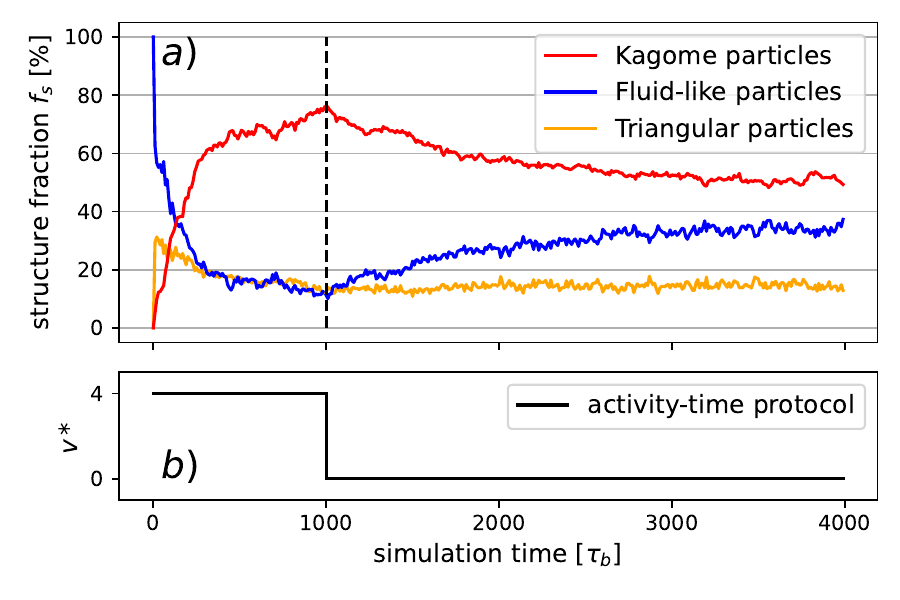}
   \end{overpic}
   \caption[]{a) Percentages of Kagome, Fluid-like and Triangular particles as a function of simulation time for an exemplary system at state point $\rho^*=0.4$ and $T^*=1.3$. The vertical dotted line indicates the time point ($t=10^3\tau_b$) where the simulation is switched from active to passive. The respective structure fractions $f_s$ relax to their passive steady-state values. b) Step-like activity-time protocol. The system is active with propulsion strength $v^*=4$ for a time range of $10^3\tau_b$. Subsequently, activity is switched off and the simulation is continued until a steady state is reached.}
   \label{fig:ATP}
\end{figure}

Motivated by the results of the previous section, we now propose an activity-time protocol, where the propulsion velocity $v^*$ changes over time. We restrict ourselves to the simplest form of such a protocol, which consists of a step function as displayed in Fig.~\ref{fig:ATP}b).
For all state points $(T^*,\rho^*)$, simulations are performed with $v^*=4$ up to the time $10^3\tau_b$. This timeframe was chosen based on the previous simulations of the purely active system, where $10^3\tau_b$ was found to be sufficient to achieve high Kagome fractions $f_s$, even close to the Kagome-Fluid boundary. After $10^3\tau_b$ the propulsion is switched off ($v^*=0$) for the remaining simulation time. As an example, we show in Fig.~\ref{fig:ATP}a) results for $\rho^*=0.4$ and $T^*=1.3$ with the described activity-time protocol.
The three curves pertain to the percentages (fractions $f_s$) of Kagome, Fluid and Triangular structures as function of time.
The time interval prior to the dashed line corresponds to the aggregation process of the active system, revealing rapidly increasing Kagome and decreasing Fluid particle fractions.
Note that the Triangular fraction peaks right after the simulation is initialized, highlighting the aforementioned intermediate nature of this structure. 

\begin{figure}[]
   \centering
   \begin{overpic}[width=.5\textwidth]{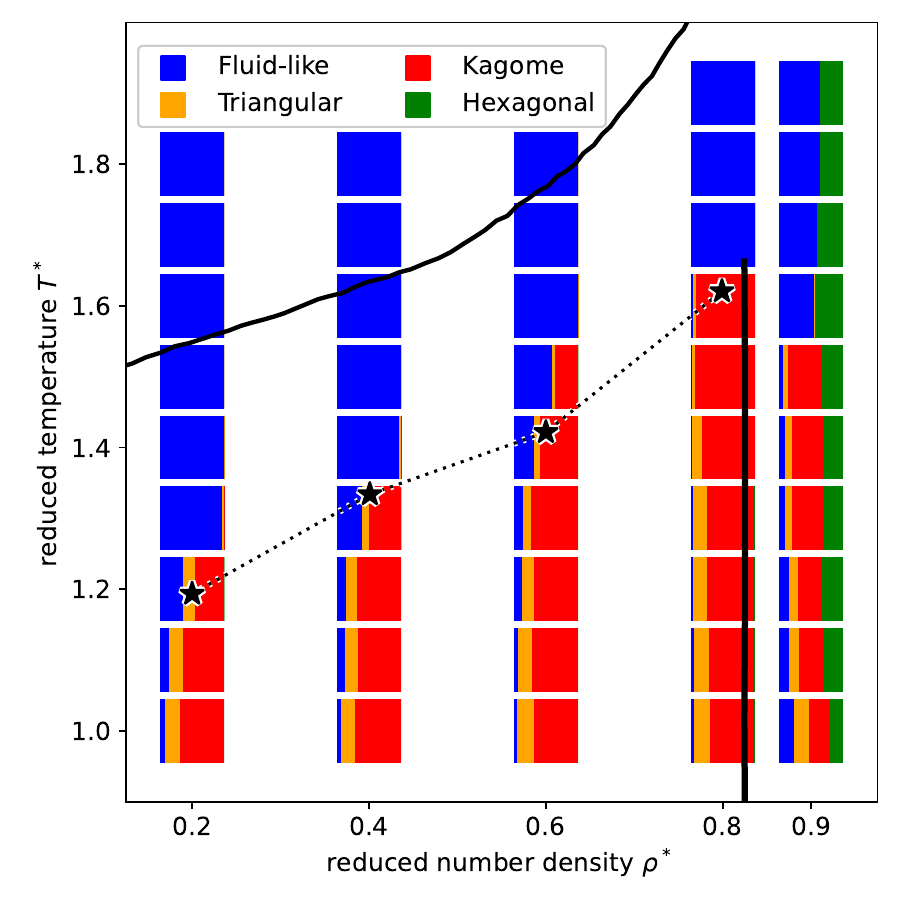}
   \end{overpic}
   \caption[]{$T^*$-$\rho^*$ state diagram with activity-time protocol sampling. Black lines and symbols indicate results from Ref.~\citenum{Sciortino}; for a detailed description, see Fig.~\ref{fig:PurelyPassiveDiagram}.}
   \label{fig:ATPSDiagram}
\end{figure}

\begin{figure*}[]
    \centering
    \fbox{\includegraphics[width=0.29\textwidth]{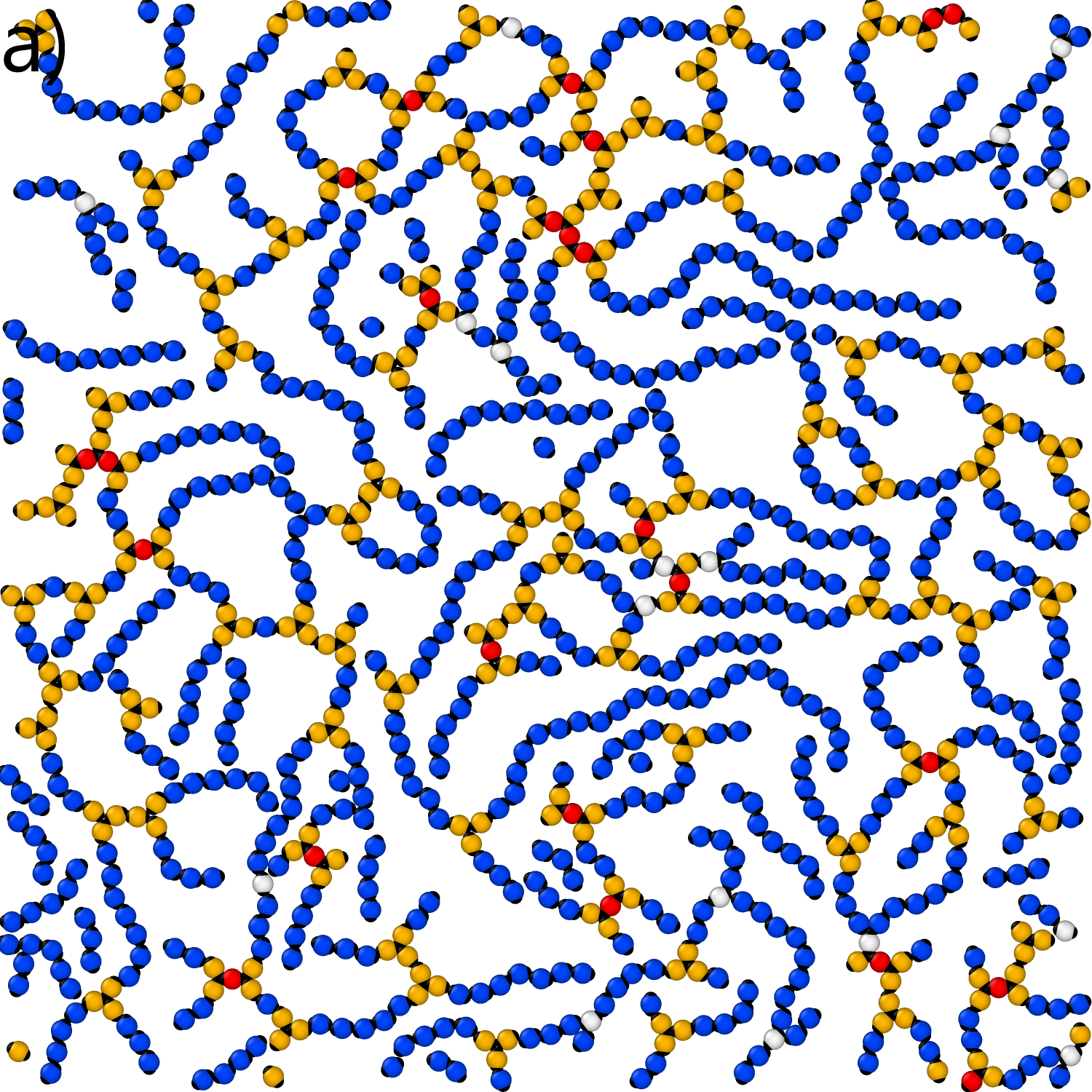}} 
    \hspace{0.02\textwidth}
    \fbox{\includegraphics[width=0.29\textwidth]{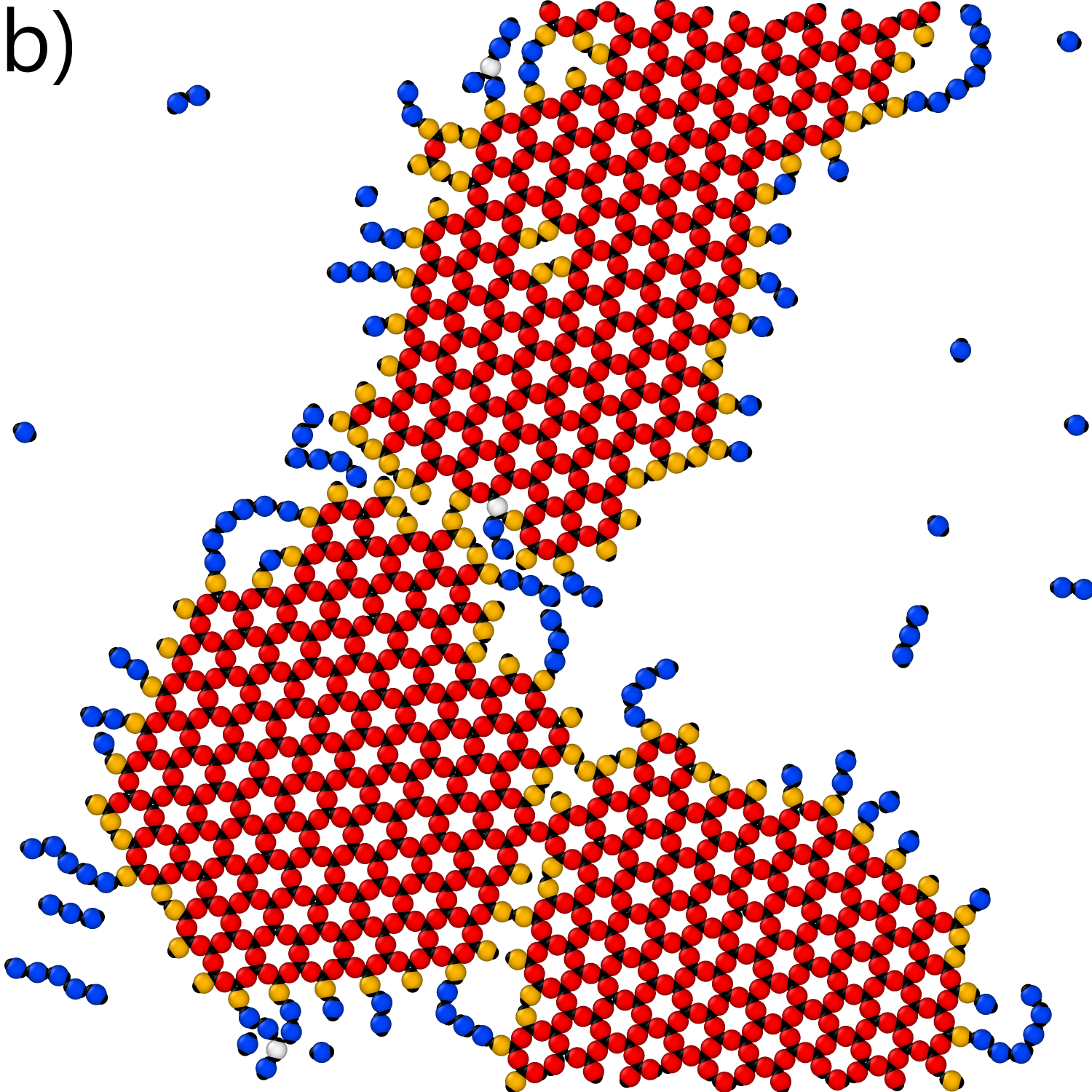}}
    \hspace{0.02\textwidth}
    \fbox{\includegraphics[width=0.29\textwidth]{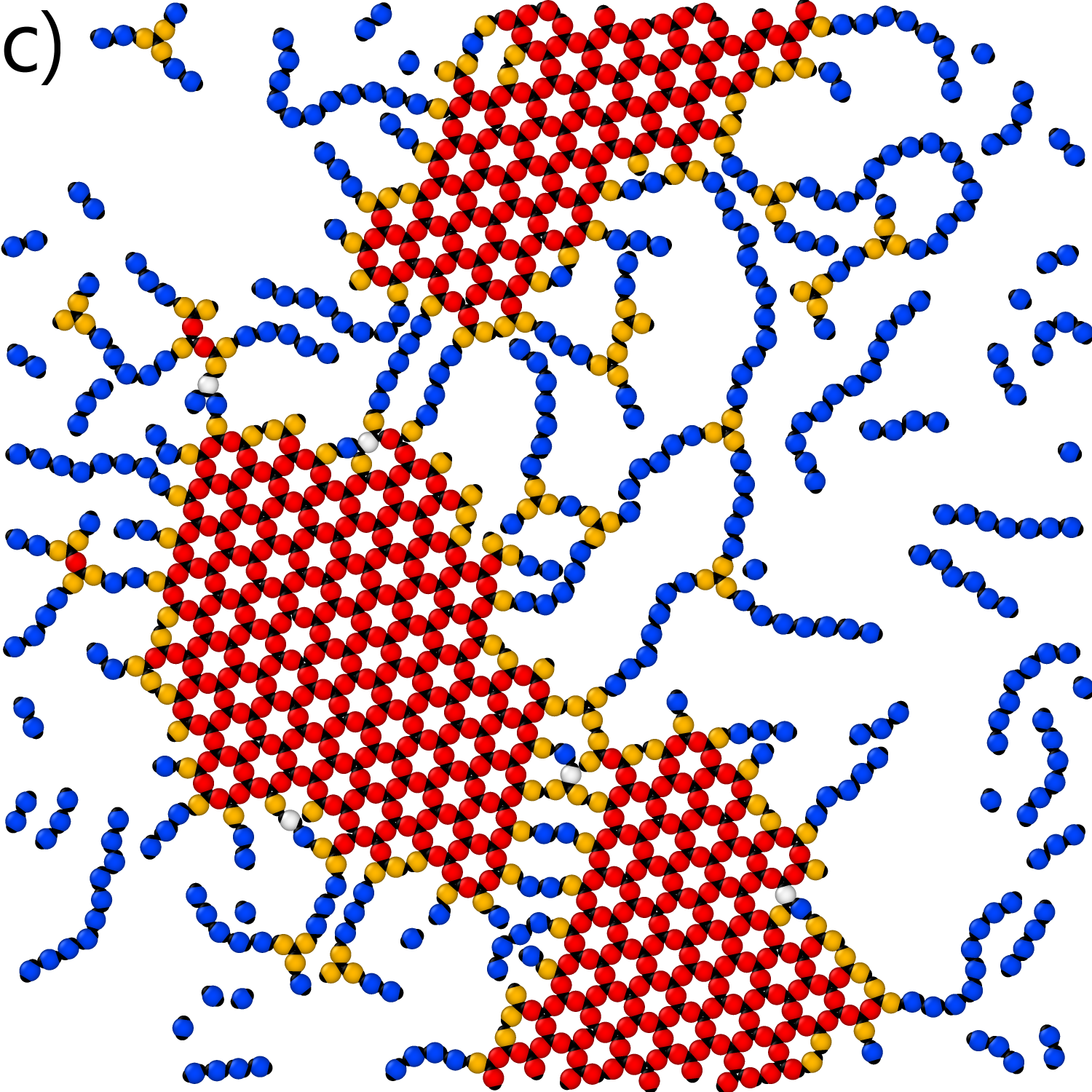}}
    \caption{Visualisation of the effect of the activity time protocol. Plots a)-c) show simulation snapshots corresponding to final configurations at $\rho^*=0.4$, $T^*=1.3$ in the $N=1024$ particle system. a) Passive case, b) purely active case and c) passive system after activity time protocol sampling. Particle colorations distinguish between Kagome (red), Triangular (orange) and Fluid-like (blue) configurations; attractive patches on each particle are indicated in black. Particle structures are determined using our detection algorithm (Appendix~\ref{sec:levelA2}).}
    \label{fig:three_images}
\end{figure*}

After switching off the activity, the assembly process seems to be partially reversed, with decreasing Kagome and increasing Fluid particle fraction. Interestingly, the Triangular fraction remains essentially unaffected by the time protocol. 
We consider the time protocol simulation as completed once the particle fractions $f_s$ have reached their steady-state values. It is important to note that the relaxation times after switching off the activity can vary significantly across different state points, ranging from a few hundred to several thousand Brownian times.
A snapshot of the final particle configuration at the end of the time protocol of an exemplary system at $\rho^*=0.4$ and $T^*=1.3$ is displayed in Fig.~\ref{fig:three_images}c).

As a side remark, we would like to mention that the process in which the Kagome crystal returns to a fluid state (either partially or completely) is a continuous and slow process, as indicated in Fig.~\ref{fig:ATP}, rather than an abrupt breaking of the cluster. The Kagome cluster is slowly dissolved from its surface once the activity is set to zero. This process is reminiscent of crystal melting upon increasing the temperature of the system \cite{ColloidalMelting}, with the difference that in our case it is the quench of activity at constant temperature that is initiating the melting process. \\

The time-averaged final configurations of the time protocol sampled state diagram are presented in Fig.~\ref{fig:ATPSDiagram}. It can be seen that a clear boundary has developed between Fluid and Kagome state points, reminiscent of the one in the active system (Fig.~\ref{fig:ActiveDiagram}). Moreover, the boundary we obtained is in close proximity to the (dotted) MC boundary of Ref.~\citenum{Sciortino}, indicating an almost quantitative agreement - now across all densities - between the two models. This is quite remarkable, considering that different models and simulation techniques were used to obtain these results.\\
Further effects of the activity-time protocol sampling can be seen from a direct comparison to the active and passive state diagrams in Fig.~\ref{fig:ActiveDiagram} and Fig.~\ref{fig:PurelyPassiveDiagram}, respectively.
Consider, for example, the state points $\rho^*=0.2$, $T^*=1.2$ and $\rho^*=0.4$, $T^*=1.3$, which exhibit Fluid-like characteristics in the purely passive system (Fig.~\ref{fig:PurelyPassiveDiagram}), and mixed Kagome-Fluid characteristics in the active system (Fig.~\ref{fig:ActiveDiagram}). For both state points, Fig.~\ref{fig:ATPSDiagram} reveals that the Kagome-Fluid coexistence persists even after activity is switched off.
A closer analysis shows that clusters formed during the initial active interval of the protocol are dissolved only partially during the subsequent relaxation process.
This is evident, exemplified for the latter state point, from Fig.~\ref{fig:ATP}a) and the snapshots presented in Fig.~\ref{fig:three_images}b) and c), which correspond to the final particle configuration in the purely active b) and activity-time protocol case c).
For other state points with reduced temperatures markably below the crystallization boundary, no noticeable change is observed between active and activity-time protocol diagrams. In such cases, the strength of inter-particle interactions seems to dominate the effects of activity.\\
We finally turn to the state points $\rho^*=0.2$, $T^*=1.3/ 1.4$ and $\rho^*=0.4$, $T^*=1.4/1.5$, which allow for the formation of a stable Kagome cluster exclusively in the active system (see Fig.~\ref{fig:ActiveDiagram}). 
Here, after switching off activity, Kagome clusters are dissolved completely, and the system evolves into a purely Fluid-like state.
We also found that direct observation of spontaneous crystallization from the homogeneous Fluid state at $\rho^*=0.2$, $T^*=1.4$ and $\rho^*=0.4$, $T^*=1.5$ is rare. This became clear from the ensemble simulations ("E" state points in Fig.~\ref{fig:ActiveDiagram}), where, in each case, only one out of ten simulations was completed with a successfully developed Kagome cluster. It is therefore instructive to study the nucleation of Kagome clusters at these state points in more detail.

\subsection{\label{sec:level8}Nucleation and cluster analysis in active triblock systems}

We have seen that activity can support the system in overcoming kinetic barriers to form a stable Kagome cluster out of a Fluid state. To gain more insight into the underlying mechanism, we here analyze the nucleation probabilities and cluster properties in the steady state.
All analyses are conducted at $\rho^* = 0.2$ and $T^* = 1.4$, where Kagome structures appear only in the active system (Fig.~\ref{fig:ActiveDiagram}) but not in the passive case (Fig.~\ref{fig:ATPSDiagram}). We explore the propulsion velocities $v^*=4$ (corresponding to Fig.~\ref{fig:ActiveDiagram}) and $v^*=8$.
In order to obtain sufficiently good statistics and reduce potential finite-size effects, we choose a system size of $N=2025$ triblock particles. For each propulsion velocity $v^*$, an ensemble of 30 independent simulation runs is performed, each for a duration of $10^3\tau_b$.

\subsubsection{\label{sec:level9}Nucleation probability}

\begin{figure}[b]
   \centering
   \begin{overpic}[width=.5\textwidth]{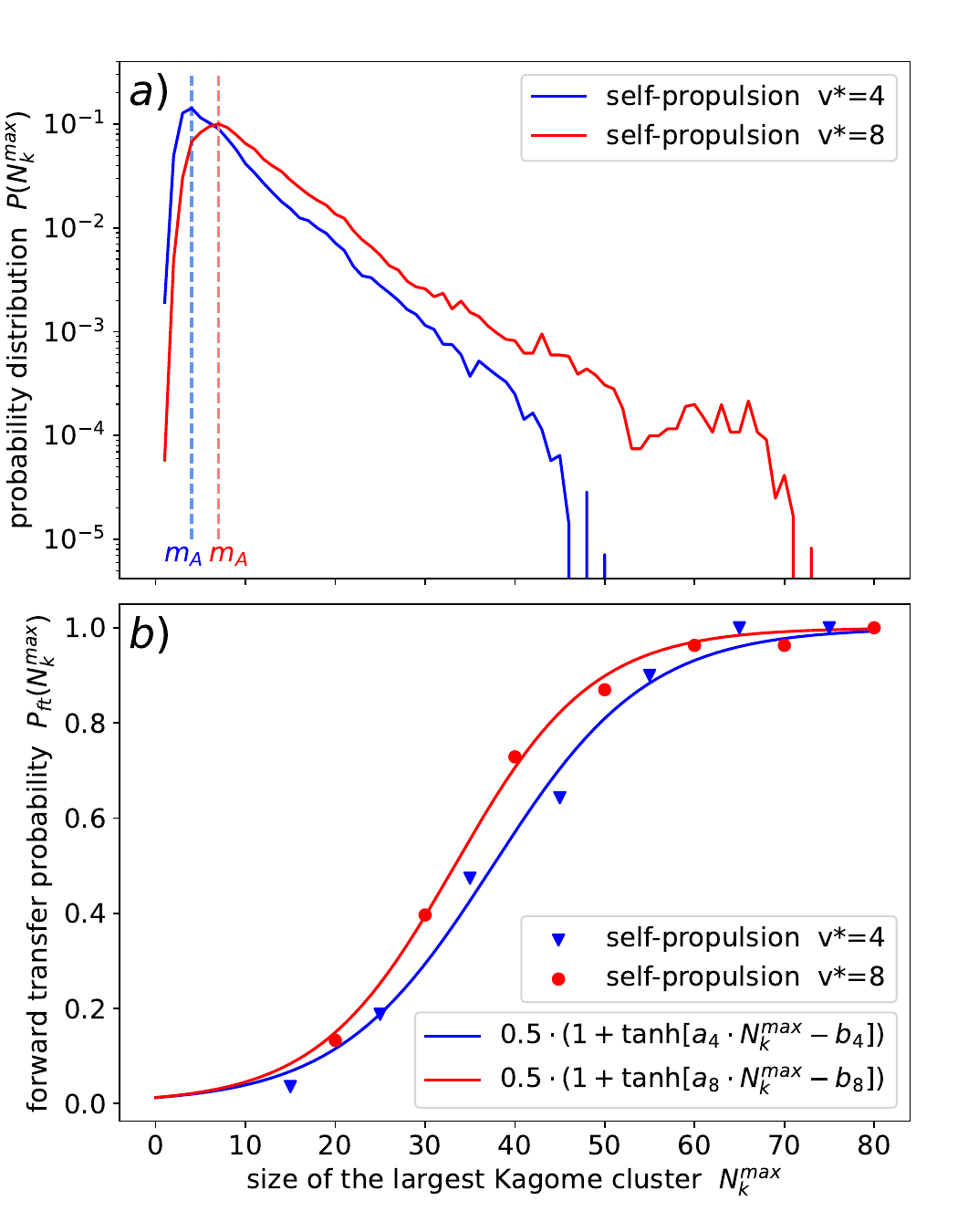}
   \end{overpic}
   \caption[]{Analysis of the formation of Kagome clusters in the active system at $\rho^*=0.2$, $T^*=1.4$ and the two self-propulsion velocities $v^*=4$ and $v^*=8$. 
   a) Probability distribution $P(N_K^{\text{max}})$ of the order parameter $N_K^{\text{max}}$ in the homogeneous state. Only configurations before a successful crystallization event contribute to the distribution. 
   b) Forward transfer probability $P_{\text{ft}}(N_K^{\text{max}})$ of a cluster of size $N_K^{\text{max}}$ to develop into a stable Kagome cluster. Blue triangles and red circles correspond to the numerical results, solid lines indicate respective fit-functions (see main text). The fit-parameters are $a_4=0.060\pm0.006$, $b_4=2.18\pm0.23$ and $a_8=0.070\pm0.004$, $b_8=2.18\pm0.13$.}
   \label{fig:Probability}
\end{figure}

To quantitatively describe the nucleation process, it is essential to define a suitable order parameter. In simple active Brownian particle (ABP) systems with isotropic interactions, the common choice for this order parameter is the size of the largest cluster.\cite{FFS}
Given our focus on the formation of Kagome clusters, a more appropriate order parameter is the size of the largest Kagome cluster, $N_K^{\text{max}}$. This quantity is determined by identifying the number of Kagome particles $N_K$ in each individual cluster and selecting the cluster with the largest Kagome fraction as $N_K^{\text{max}}$.\\
Crystallization at the selected state point occurs spontaneously from a homogeneous Fluid-like state. 
This process is initiated by the formation of small, localized Kagome clusters, accompanied by strong fluctuations in $N_K^{\text{max}}$. These clusters may either grow, leading to phase separation, or dissolve, returning the system to its homogeneous state. 
In this context, an important quantity is the size of the critical nucleus, $N_{\text{crit}}$. Only clusters with a size larger than $N_{\text{crit}}$ will grow.\\
Following the rare event analysis of Richard \textit{et al}. \cite{FFS}, we here define $N_{\text{crit}}$ as the average value of $N_K^{\text{max}}$ at which the probability of reaching the phase-separated state is $P_{\text{ft}}(N_{\text{crit}})=0.5$.
Here, $P_{\text{ft}}$ represents the forward transfer probability, which denotes the likelihood that a Kagome cluster of size $N_K^{\text{max}}$ develops the system into a phase-separated state. This probability is determined using the Direct Forward Flux Sampling method (see in Ref.~\citenum{DFFS} and Appendix~\ref{sec:DFFS}).
As a first step, we generate the probability distribution $P(N_K^{\text{max}})$ of the fluctuating order parameter $N_K^{\text{max}}$ within the homogeneous Fluid-like state. 
Results are presented in Fig.~\ref{fig:Probability}a) for both propulsion velocities. 
For each distribution, the value $m_A$ related to the location of the maximum represents the mean of $N_K^{\text{max}}$. We note that, on a microscopic level, the term "homogeneous state" is justified only in the regime of $P(N_K^{\text{max}})$ for $N_K^{\text{max}}<m_A$.
The other regime, where $N_K^{\text{max}}>m_A$ corresponds to a Fluid-like state containing small and temporally unstable Kagome nuclei. Still, we can consider the entire state underlying the data in Fig.~\ref{fig:Probability}a) as homogeneous on average, since the corresponding local density distribution shown in Fig.~\ref{fig:Cluster_Stability}a) exhibits only one, weakly pronounced maximum around the average density, $\rho^*=0.2$. Details on the calculation of the local density distribution are given in Appendix~\ref{sec:levelA3}. In the subsequent Appendix~\ref{sec:DFFS} we describe the application of Direct Forward Flux Sampling to our system.\\
Our results for the forward transfer probability $P_{\text{ft}}(N_K^{\text{max}})$ at both propulsion velocities, are shown in Fig.~\ref{fig:Probability}b). The numerical data are represented by blue triangles ($v^*=4$) and red circles ($v^*=8$), respectively. We have fitted the data with a fit function of the form $0.5\cdot(1+\tanh[a\cdot N_K^{\text{max}}+b])$, as previously used in Ref.~\citenum{FFS} for an ABP system in the regime of motility-induced phase separation\cite{DoublePeak} (MIPS) close to the binodal. It is seen that this function provides indeed a reasonable fit, as indicated by the solid curves in Fig.~\ref{fig:Probability}b). 
This already indicates that nucleation of Kagome clusters in active triblock systems shares some similarities with MIPS-driven cluster formation in ABP systems.
With the fit function, the critical nucleus size $N_{\text{crit}}$, following from the condition $P_{\text{ft}}(N_{\text{crit}})=0.5$, is given by $N_{\text{crit}}=a/b$.
We find $N_{\text{crit}}(v^*=4)=38\pm8$ and $N_{\text{crit}}(v^*=8)=33\pm4$. Thus, the values of $N_\text{crit}$ are approximately equal at the two propulsion velocities considered, with a tendency towards a lower critical nucleus size in the $v^*=8$ system. Here, we also observe consistently higher probabilities $P(N_K^{\text{max}})$. Moreover, only at $v^*=8$ the probability distribution extends towards high values $N_K^{\text{max}}$ where $P_{\text{ft}}(N_K^{\text{max}})$ approaches unity.
At first glance, the propulsion velocity $v^*=8$ thus appears more suitable for sampling the Kagome configuration and should therefore also be more suitable for application of our activity time protocol. However, our observations (data not shown) indicate that the majority of simulated $v^*=8$ systems do not reach a stable configuration consisting of a single large Kagome cluster. Instead, these systems tend to undergo partial compactification into hexagonal structures or even fragment into multiple smaller clusters. For the subsequent analysis of the steady state properties, we therefore consider only the case $v^*=4$.

\subsubsection{\label{sec:level10}Steady-state cluster analysis}

\begin{figure}[b]
   \centering
   \begin{overpic}[width=.5\textwidth]{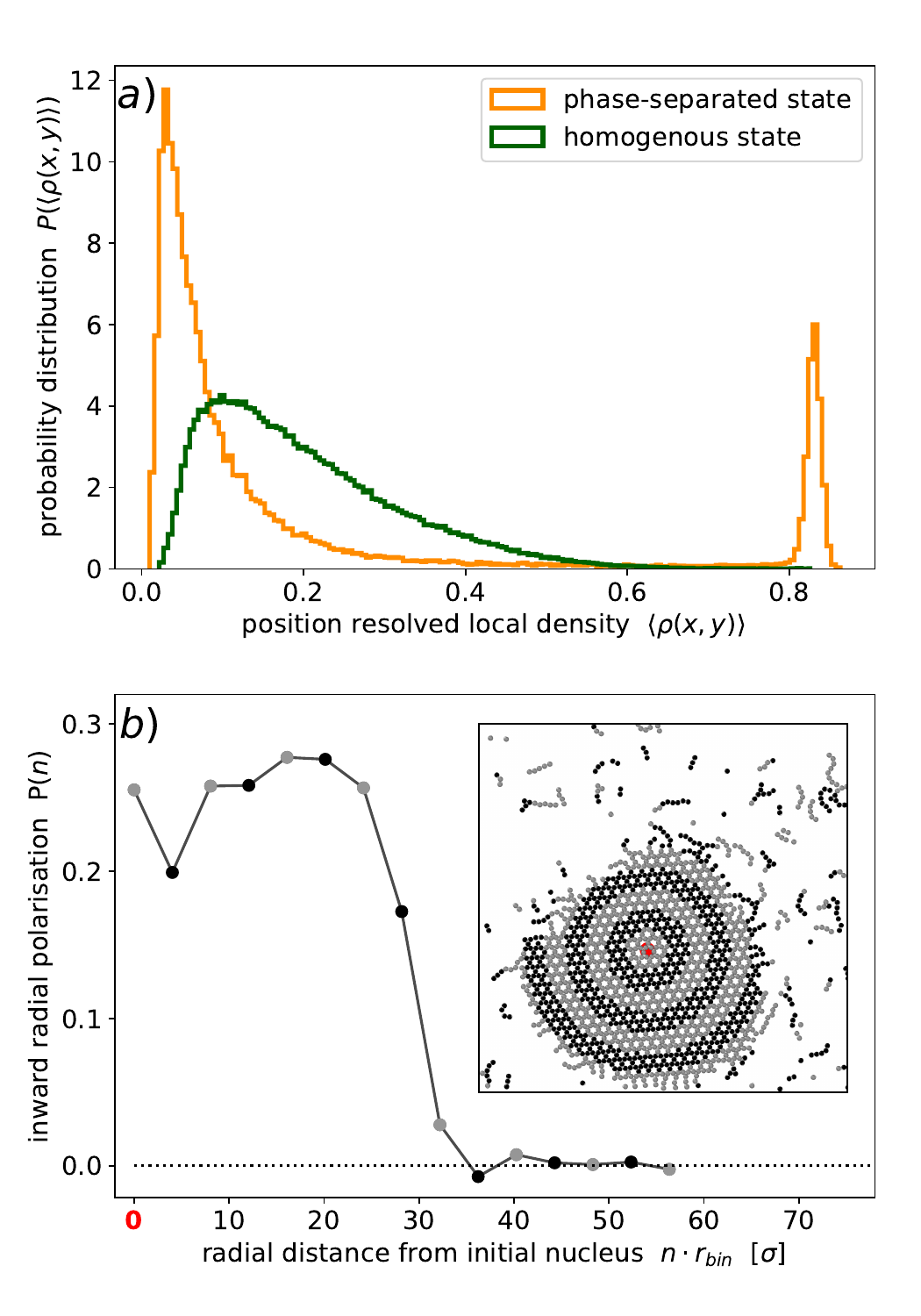}
   \end{overpic}
   \caption[]{Analysis of the steady state at $\rho^*=0.2$, $T^*=1.4$, $v^*=4$. a) Probability distribution of the position-resolved local density $\langle \rho(x,y) \rangle$. The homogeneous state distribution (green) is calculated before the onset of crystallization. The final phase-separated state is characterized by a double-peak distribution (orange).
   b) Radial polarization as a function of bin index $n$ and zoomed-in simulation snapshot of an exemplary Kagome cluster with a surrounding dilute phase as an inset. The position of the initial nucleus is marked in red, with its surrounding bins highlighted via alternating black and grey tones. The corresponding markers of $\text{P}(n)$ indicate our numerical results on the local polarization within each bin $n$. The solid line is drawn as a guide to the eye.}
   \label{fig:Cluster_Stability}
\end{figure}

To investigate the structures in the steady state at $\rho^*=0.2$, $T^*=1.4$, $v^*=4$, we first consider the position-resolved local density distribution, $P(\langle \rho(x,y) \rangle)$, obtained as described in previous studies \cite{starkMIPS, LiaoMIPS} (see also Appendix~\ref{sec:levelA3}). This analysis is performed both, before and after the onset of crystallization.
Results are shown in Fig.~\ref{fig:Cluster_Stability}a), where the green curve represents the homogeneous state before crystallization, and the orange curve corresponds to the phase-separated state after successful crystallization, i.e., when the Kagome cluster has reached its steady-state configuration. In the first case we observe a broad distribution with one weakly pronounced maximum typical of a, on average, homogeneous state. In contrast, in the second, phase-separated case a characteristic double-peak structure emerges, resembling the bimodal density distributions observed in the MIPS regime of ABP systems \cite{DoublePeak}. Further insights can be gained by analyzing the local particle polarization within the cluster.\\
To this end we compute the net alignment of the particles' heading vectors, $\hat{\pmb{n}}$, relative to the cluster center. The latter is defined as the mean position of the first ($1<N<10$) particles that formed the initial nucleus, onto which additional particles adsorb during cluster growth. Note that this center of reference does not necessarily coincide with the geometric center of the cluster at later times.
To quantify the local polarization within the cluster, the simulation box is divided into concentric circular bins of width $r_{\text{bin}} = 4\sigma$, extending radially outward from the cluster center, as illustrated in the inset of Fig.~\ref{fig:Cluster_Stability}b). Each annular bin, indexed by $n$, is confined between the radii $n\cdot r_{\text{bin}}$ and $(n+1)\cdot r_{\text{bin}}$. The polarization within a given bin is then defined as

\begin{equation} 
\text{P}(n) = \biggl\langle \frac{1}{N_n}\sum_{i \in n}^{N_n} \hat{\pmb{n}}_i\cdot \hat{\pmb{r}}_{ic} \biggr\rangle_{t,e} \quad . \label{eq:pol} 
\end{equation}

Here, the summation extends over all $N_n$ particles within the $n$-th bin. For each bin, we compute the dot product between the particle's unit heading vector $\hat{\pmb{n}}_i$ and the radial unit vector $\hat{\pmb{r}}_{ic}$, that points from particle $i$ to the cluster center. The notation $\langle.\rangle_{t,e}$ denotes an average over a time window of $100\tau_b$ in the steady state, as well as an ensemble average over five independent simulations. This broad time interval is necessary to ensure that polarization averages to zero in the dilute regime. Results for the polarization as a function of distance from the initial nucleus are presented in Fig.~\ref{fig:Cluster_Stability}b).
We clearly observe a nonzero polarization, directed inwards, for all distances within the cluster, which extends radially up to approximately $30\sigma$ from the initial cluster nucleus. Beyond the cluster region, the polarization rapidly drops to zero.
The emergence of a nonzero polarization within the Kagome cluster of our active triblock system reminds of that in ABP systems close to interfaces. Indeed, in ABP systems, polarization is commonly observed near solid boundaries \cite{SolidBoundary_CLInterface}, in active-passive interfaces \cite{APInterface1, APInterface2}, and in regions where dense and dilute phases coexist due to motility-induced phase separation (MIPS) \cite{ClusterInterface1, ClusterInterface2, SolidBoundary_CLInterface}. In the context of the clusters accompanying MIPS, the polarization typically results from self-propelling particles accumulating at the cluster boundary, pointing toward the dense phase. Inside such ABP clusters, rotational diffusion randomizes particle orientations on the order of the rotational timescale $\tau_r$. This leads to a vanishing net polarization within the inner ABP cluster, leaving only a polarized interfacial layer.
In contrast, in the active triblock system studied here, the bonds formed between attractive patches strongly restrict rotational freedom. Consequently, the Kagome structures that emerge at the interface during early cluster formation retain their interfacial polarization; effectively, their orientation is frozen-in.\\
The excess of self-propelled and inward-pointing particles generates a force, exerting pressure toward the cluster center. This inward-pointing force likely contributes to the stability of the Kagome clusters in the $v^*=4$ system. Moreover, the effect may also explain the partial collapse of Kagome clusters at $v^*=8$. Due to the stronger self-propulsion, the resulting inward force is even enhanced, leading to the compactification of Kagome structures into Hexagonal ones.\\
The emergence of an inward polarization in active triblock clusters was also discussed in Ref.~\citenum{mallory2019}. However, in that study, polarization was only reported in Hexagonal clusters for systems at lower reduced temperature and intermediate density, where activity primarily serves to enhance structure formation rather than to induce it. Here, we have deliberately studied a state directly at the crystallization boundary, where activity is essential for successful crystallization and cluster stability, revealing a new type of a radially polarized Kagome cluster.

\section{\label{sec:Conclusion}Conclusion}

In this study, we performed extensive BD simulations of a model system\cite{mallory2019} of "patchy" (triblock) Janus colloids to systematically explore the state space in equilibrium (passive case), as well as in presence of activity. Our categorization of states is based on the analysis of local structures for which we have suggested an automatized algorithm.
A particular focus has been on the range of densities and temperatures where, in earlier MC simulations of a related (passive) Janus colloid model\cite{Sciortino}, spontaneous formation of crystalline Kagome clusters from the fluid phase was observed in a broad range of densities.

Our present simulations in the purely passive case revealed the onset of Kagome crystallization at high densities $\rho^*=0.6$-$0.8$, in quantitative agreement with Ref.~\citenum{Sciortino}. 
However, at lower densities ($\rho^* = 0.2$-$0.4$), we rather observed the emergence of a web-like network of particles which persisted on the time scale of the simulations (similar to the findings in other studies\cite{Granick, mallory2019}), corresponding to a metastable state.

To overcome these kinetic limitations, we here proposed a sampling involving an "activity-time protocol". This idea was inspired by an earlier BD study\cite{mallory2019} showing that an artificially introduced self-propulsion along the particle's symmetry axis can strongly support the self-assembly into Kagome clusters, depending on the motility $v^*$ that was a fixed parameter in Ref.~\citenum{mallory2019}.
In our present study, we employ a time-dependent activity $v^*(t)$. The simulations are started and conducted with a constant value $v^*>0$ until the resulting active system approaches a nonequilibrium steady state characterized by a large fraction of Kagome-coordinated particles.
We then "quench" the system into its passive limit characterized by $v^* = 0$, and allow for equilibration. The state diagram obtained with this step-like activity protocol
reproduces the low-density crystallization boundary of the equilibrium MC simulations in Ref.~\citenum{Sciortino}, showing that the time protocol can indeed prevent a trapping into metastable networks.

To get more physical insight into the impact of activity at low densities and temperatures close to the boundary, we conducted a Direct Forward Flux Sampling analysis to calculate the nucleation probability from the homogeneous fluid, using the size of the largest Kagome cluster as an order parameter.  The results reveal interesting analogies to the crystallization of equilibrium Lennard-Jones systems\cite{LJcrystallization},
as well as to systems of active Brownian particles (ABP) close to motility-induced phase separation\cite{FFS}. We also found analogies to the ABP system in the steady state achieved after cluster formation: Here we
observe a bimodal density distribution indicating a phase-separated state, and an inward-pointing polarization of particles within the Kagome cluster. An important difference to ABP clusters, however, 
is that in our system, the polarization extends into the inner part of the cluster, contrary to the interfacial polarization effects seen in ABP clusters. 
This difference results from the presence of strongly anisotropic pair interactions between two patchy, active Janus particles. 
These interactions restrict rotational diffusion and effectively "lock in" the initial surface polarization acquired during previous stages of cluster formation. Moreover, the polarization seems to have an important effect on the cluster stability:
While we found stable Kagome clusters at $v^* = 4$, the clusters collapsed at $v^* = 8$, although the likelihood of formation of a Kagome cluster is larger in the latter case. This indicates a subtle interplay of the model ingredients for the formation of stable Kagome clusters. In view of these findings, it would be intriguing to design an activity-time protocol, which initially sets the system at $v^* = 8$ to enhance crystallization, followed by a controlled reduction of the propulsion velocity to $v^* = 4$ before any cluster collapse occurs. Such a purely active protocol could further improve the sampling of the $v^* = 4$ state diagram in the vicinity of the crystallization boundary. To implement this approach effectively, an extended nucleation analysis would be necessary, particularly at higher densities. More generally, an issue that should be explored in more detail concerns the choice of switching time between the time ranges of constant $v^*$. Finally, while we have considered here a step-wise protocol, one could try to optimize the function $v^*(t)$ similar to efforts in other areas of control of nonequilibrium soft matter. \cite{Golestanian, AIProtocol, Hagan, BiskerControl}\\ 
A key advantage of employing such activity protocols is their potential applicability in experimental settings. 
The possibility of time-dependent control of self-propulsion has already been demonstrated in experimental setups for light-induced active Janus particles without hydrophobic (attractive) patches upon controlling the light intensity.\cite{LightInducedReverse, ExpActiveDoping}
To our knowledge, no experimental realization of active triblock Janus particles exists to date. 
Nevertheless, the synthesis of both passive triblock Janus particles \cite{Granick, GLAD} and (single patch) active Janus particles\cite{ExpActiveAmphiphilic} has been demonstrated, where Ref.~\citenum{ExpActiveAmphiphilic} used a chemical propulsion mechanism\cite{ChemPropulsion} in combination with hydrophobic interactions (analogous to Ref.~\citenum{Granick}).
The fabrication of an active triblock Janus particle seems therefore feasible. 
We also hope that the here presented approach of using an activity-time protocol is applicable also to other colloidal systems with strongly anisotropic interactions that tend to stick in arrested states such as patchy particles\cite{PatchyColloidReview} and colloidal gels\cite{ZaccarelliReview}, or dipolar\cite{LiaoDipolar, ArrestedGels} and multipolar\cite{KoglerMultidirectional} particles.

\begin{acknowledgments}
This work was funded by the Deutsche Forschungsgemeinschaft (DFG, German Research Foundation), project number 449485571.
\end{acknowledgments}

\appendix

\section{\label{sec:levelA1}Relation to the Kern-Frenkel model}

In this Appendix we discuss relations between the present model, Eq.~(\ref{eq:potential}), and the Kern-Frenkel potential used in Ref.~\citenum{Sciortino}.
The Kern-Frenkel (KF) potential\cite{KFpotential} can be written as a product of the isotropic square-well potential $\Psi_{\text{SW}}(r_{ij})$, and an anisotropic angular potential, $\phi_{\text{KF}}(\theta_i, \theta_j)$.
The square-well potential is defined as 

\begin{equation}
    \Psi_{\text{SW}}(r_{ij}) = 
    \begin{cases}
    \infty  & , \quad r_{ij} < \sigma\\
    -u_0 & , \quad \sigma < r_{ij} < (1+\delta)\sigma \\
    0 & , \quad r_{ij} > (1+\delta)\sigma \quad ,
\end{cases}
\end{equation}

where in Ref.~\citenum{Sciortino}, the attractive range is chosen as $\delta=0.05$ and the potential energy well depth is set by $u_0$. The angular potential is chosen as a step function, that is,

\begin{equation}
    \phi_{\text{KF}}(\theta_i, \theta_j) = 
    \begin{cases}
    1  & , \quad \theta_i < \theta_{\text{KF}} \ \text{and} \ \theta_j < \theta_{\text{KF}}\\
    0 & , \quad \text{otherwise}  \label{eq:KFangular} \quad .
\end{cases}
\end{equation}

Thus, $\phi_{\text{KF}}=1$ if both angles $\theta_i$ and $\theta_j$ are smaller than $\theta_{\text{KF}}$ and zero otherwise. The definition of $\theta_i$ and $\theta_j$ is the same as in section~\ref{sec:level2}. Note that $\phi_{\text{KF}}(\theta_i, \theta_j)$ can also be understood as the limiting case of the product potential $\phi(\theta_i)\phi(\theta_j)$, appearing in Eq.~(\ref{eq:potential}), in the limit $\theta_{\text{tail}} \to 0$, with $\theta_{KF}$ playing the role of $\theta_{\text{max}}$. With the present choice $\theta_{\text{tail}}=25^\circ$, there is a broad range of nonzero values of the angular potential above $\theta_{\text{max}}$, as can be seen from the inset of Fig.~\ref{fig:DistancePot}. This will generally lead to a relation $\theta_{\text{max}}<\theta_{\text{KF}}$, where $\theta_{\text{max}}$ in our model needs to have smaller values than $\theta_{KF}$ in order to account for the $\theta_{\text{tail}}$ region.
To find a suitable value of $\theta_{\text{max}}$ in the present model, we have visually compared simulation snapshots of our simulations and a supplemental video provided in Ref.~\citenum{Sciortino} at $\rho^*=0.6$, $T^*\approx 1.4$. This has led to the choice $\theta_{\text{max}}=26^\circ$, which is the smallest angle allowing for the formation of Kagome structures at the respective temperature and density. Our value for $\theta_{\text{max}}$ is close to the choice $\theta_{\text{max}}=25.2^\circ$ in Ref.~\citenum{mallory2019} which, however, does not allow for the formation of Kagome structures at this state point. \\
We now turn to the coupling strength measured by the dimensionless temperature.
In Ref.~\citenum{Sciortino}, the reduced temperature for the KF potential is defined as $T^*_{\text{SC}}=k_BT/u_0$, which relates the depth of the potential well to the inverse of the reduced temperature.
The reduced temperature in our model is given by $T^*=k_BT/\epsilon_{\text{rep}}=15 k_BT/\epsilon_{\text{att}}$, which can not directly be related to the potential well depth of $U^*_{ij}$. We treat this problem as follows.
By locating the minimum $r^*_{\text{min}}$ of our reduced potential $U^*_{ij}=U_{ij}/k_BT$ we first establish the relation $U^*_{ij}(r^*_{\text{min}}) = \epsilon_{\text{eff}} \approx -11.33/T^*$ between our definition of $T^*$ and the well depth $\epsilon_{\text{eff}}$ of our potential.
In order to find a relation between the reduced temperatures of both models, we then match the potential well depths by setting $u_0=\epsilon_{\text{eff}}$, resulting in $T^* \approx 11.33 \cdot T^*_{\text{SC}}$. The close proximity of the crystallization boundaries observed in our model and that in Ref.~\citenum{Sciortino}, as presented in Fig.~\ref{fig:ATPSDiagram}, suggests that this approach is indeed successful in translating between the reduced temperatures.

\section{\label{sec:levelA2}Structure detection algorithm}

\begin{figure*}[]
   \centering
   \begin{overpic}[width=.9\textwidth]{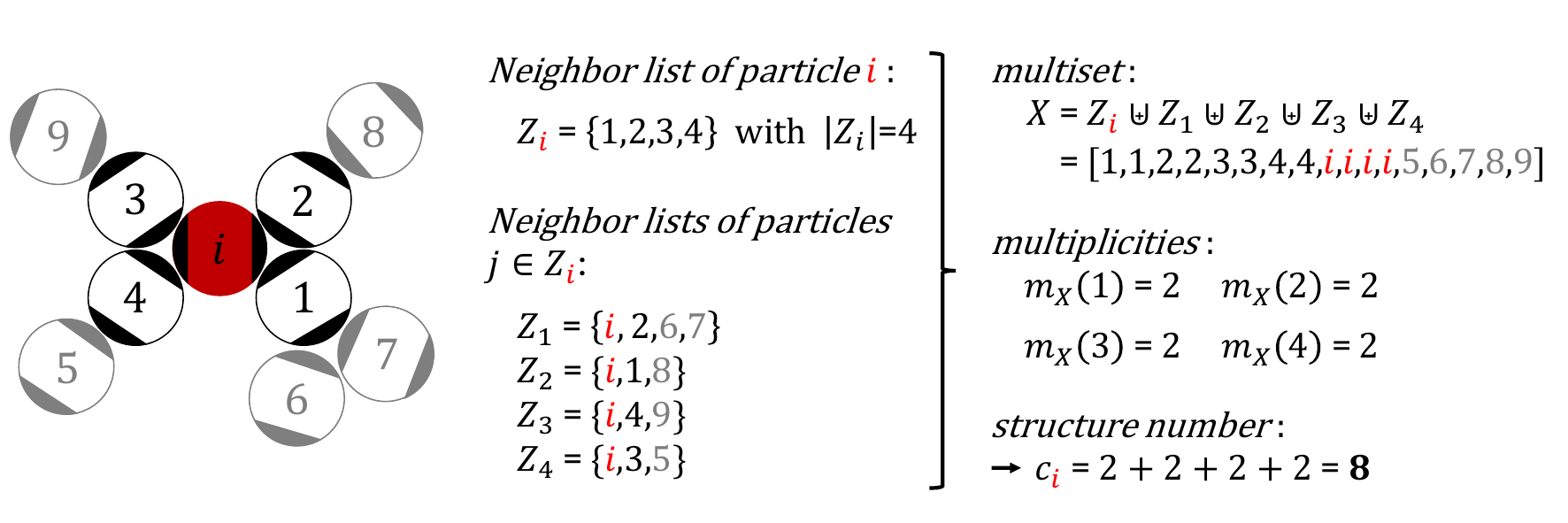}
   \end{overpic}
   \caption[]{Sketch of a particle $i$ in a Kagome configuration. The particle is surrounded by its immediate neighbors $j$ (black), that is, $i$ and $j$ are bonded. Also shown is an exemplary configuration of next nearest neighbors (grey). All surrounding particles are labeled with numbers to demonstrate the calculation of the unique Kagome structure number $c_i$ from the multiset $X$ and its multiplicities $m_X(j)$, as explained in the text.}
   \label{fig:KagomeScetch}
\end{figure*}

\begin{table}[b]
\caption{Definition of structure types in our algorithm.}
\begin{tabular}{|c|c|c|c|c|c|c|}
\hline
            & \begin{tabular}[c]{@{}c@{}}a) \\ Hexagonal\end{tabular} & \begin{tabular}[c]{@{}c@{}}b) \\ Kagome\end{tabular} & \begin{tabular}[c]{@{}c@{}}c1) \\ Triangular\end{tabular} & \begin{tabular}[c]{@{}c@{}}c2) \\ Triangular\end{tabular} & \begin{tabular}[c]{@{}c@{}}d) \\ Chain\end{tabular} & \begin{tabular}[c]{@{}c@{}}e) \\ Monomer\end{tabular} \\ \hline
$c_i$   & 4            & 8         & 4              & 5              & 2        & -          \\ \hline
$|Z_i|$     & 4            & 4         & 2              & 3              & 2        & 0          \\ \hline
$z_i$ & 6            & -         & -              & -              & -        & -          \\ \hline
\end{tabular} \label{tab:structures}
\end{table}

Here we describe our strategy to identify relevant structures emerging during the self-assembly process. The first step consists of identifying all pair bonds at a given time. Two particles $i$ and $j$ are considered bonded (i) if their distance satisfies the condition $r_{ij}<1.15\sigma$, and (ii) if an attractive interaction exists between any two of their patches, fulfilling the condition $\phi(\theta_i)\phi(\theta_j)>0$. The latter criterion is particularly important at high densities where a pure distance-based criterion would be insufficient. For an exemplary configuration, see Fig.~\ref{fig:KagomeScetch}.

All neighboring particles $j$ that are bonded to particle $i$ are stored in a neighbor list
\begin{equation}
    Z_i = \{j \ \ | \ \ r_{ij}<1.15\sigma \ \ \text{and} \ \ \phi(\theta_i)\phi(\theta_j)>0 \} \quad .
\end{equation}

Henceforth, we treat $Z_i$ as a mathematical set whose cardinality $|Z_i|$ describes the number of bonds associated with particle $i$ (see Fig.~\ref{fig:KagomeScetch} left for an example). Further, we use $Z_i$ as a basis to construct a multiset $X$, allowing us to infer a unique structure number. The multiset is defined by Eq.~(\ref{eq:multiset}) according to:
\begin{equation}
    X = Z_i \ \uplus \ Z_k \quad \text{with} \quad Z_k=\biguplus_{j \in Z_i} Z_j \quad . \label{eq:multiset}
\end{equation}

Here, $\uplus$ denotes multiset addition \cite{Blizard, Multisets}, which allows elements to appear multiple times in $X$ and $Z_k$, unlike union operations in standard set theory. 
$Z_k$, in this context, is the multiset generated from summing neighbor lists $Z_j$ of particles $j \in Z_i$. 
The number of occurrences of an element $j$ in $X$ is referred to as its multiplicity $m_X(j)$, as illustrated by the example in Fig.~\ref{fig:KagomeScetch} (right). \\
We are now in the position to create an integer number $c_i$, unique to each structure type. This is done by summing up the multiplicities $m_X(j)$ of elements $j \in X$, which are also contained in $Z_i$, according to
\begin{equation}
    c_i=\sum_{j \in Z_i}m_X(j) \quad .
\end{equation}

The so obtained integer number $c_i$ for each structure type is given in Table~\ref{tab:structures}.
The combination of $c_i$ and the number of bonds $|Z_i|$ is sufficient to uniquely classify the structure type of particle $i$, according to one of the structures presented in Fig.~\ref{fig:structuretypes}.\\
Some additional care is taken for hexagonal structures. Here we use as an additional criterion the coordination number $z_i$, i.e., the number of nearest neighbors (obtained on the basis of the distance criterion $r_{ij}<1.15\sigma$ alone). This refinement improves the detection accuracy at hexagonal cluster interfaces. The origin of the correction arises from the fact that the top and bottom particles in configuration a) in Fig.~\ref{fig:structuretypes} are not bonded to the central hexagonal particle via patch interaction.\\
The described multiset analysis can be implemented in \textit{Python} as
\begin{equation}
    c_i = \text{len}(j \ \ \text{for} \ \ j \ \ \text{in} \ \ X \ \ \text{if} \ \ j \ \ \text{in} \ \ Z_i) \quad .
\end{equation}

Finally, we note from Fig.~\ref{fig:KagomeScetch} that the configuration of next nearest neighbors (grey particles) around the central Kagome configuration ($i$ + black particles) is independent of its unique structure number $c_i$. Consequently, the detection algorithm works in both, crowded and dilute, environments.
Indeed, we expect that our algorithm will be generally applicable for structure detection in other patchy particle systems or other particle systems with strong anisotropic interactions and resulting favoured configurations.

\section{\label{sec:levelA3}Position-resolved local density}

The position-resolved local density distributions in Fig.~\ref{fig:Cluster_Stability}a) were obtained by first calculating the particle-resolved local densities $\rho_i=1/A_i$. Here, the area $A_i$ around each particle $i$ is obtained by performing a Voronoi tessellation\cite{Voronoi} of the simulation box. To account for the periodic boundaries, eight identical images of the main simulation box are generated and positioned around it, such that Voronoi cells close to the boundaries are correctly partitioned.\cite{LiaoMIPS} Local densities are calculated exclusively for particles within the main simulation box. Based on this information, we can calculate the position-resolved local densities $\rho(x,y)$, where $(x,y)$ are the coordinates of a square-like pixel with area $A_{\text{px}}\approx\sigma^2$. The density $\rho(x,y)$ is defined as \cite{starkMIPS}
\begin{equation}
    \rho(x,y) = \sum_{k} \rho_k \frac{A_{\text{ov}}(k)}{A_{\text{px}}} \quad , \label{eq:PosResolved}
\end{equation}

where we sum over the densities $\rho_k$ belonging to all Voronoi cells $k$ that overlap with the considered pixel.
This sum is weighted by the area fraction of each Voronoi cell's overlap area $A_{\text{ov}}(k)$ with the pixel, normalized by the pixel area $A_{\text{px}}$. A time average of $20\tau_b$ is performed on each pixel density $\langle \rho(x,y) \rangle$ to average over transient structures in the dilute regime of the simulation box. Within this time interval the position of a cluster, if existent, remains approximately unchanged.
The distribution of all averaged pixel densities $\rho(x,y)$ is presented in the histograms in Fig.~\ref{fig:Cluster_Stability}a).

\section{\label{sec:DFFS}Direct Forward Flux Sampling analysis}

To perform a Direct Forward Flux Sampling\cite{DFFS} analysis, a series of $M$ interfaces $m_i=\{ m_A, m_1, ...,m_B \}$ is defined at specific order parameter values $N_K^{\text{max}}=m_i$. From this series, the conditional probabilities $P(m_{i+1}|m_i)$ for transitions between interfaces $m_i\to m_{i+1}$ are obtained. The final interface $m_B$ represents the phase-separated state and is selected such that $P(m_{B}|m_{B-1})=1$. For the studied systems, suitable interface selections were identified as $m_i(v^*=4)=\{15,25,...,75 \}$ and $m_i(v^*=8)=\{20,30,...,80 \}$, where it is essential to choose interfaces with sufficiently large separations to ensure that the transition probabilities $P(m_{i+1}|m_i)$ remain approximately independent.\\
From an ensemble of 30 simulations, the conditional probability $P(m_{i+1}|m_i)$ is determined by calculating the fraction of configurations at interface $m_i$ that successfully reach interface $m_{i+1}$ relative to those that return to interface $m_A$. The forward transfer probability can then be computed from these conditional probabilities as
\begin{equation}
    P_{\text{ft}}(m_i) = \prod_{j=i}^{M-1} P(m_{j+1}|m_j) \quad .
\end{equation}

\nocite{*}
\bibliography{aipsamp}

\end{document}